\documentclass{aa}

\usepackage{color}

\usepackage{amsmath}
\usepackage{natbib}
\usepackage{graphicx}
\bibpunct{(}{)}{;}{a}{}{,}

\begin{document}
   \title{Scattering line polarization in rotating, optically thick disks}

   \subtitle{}

   \author{I. Mili\'{c}\inst{1,2}
   \and
   M. Faurobert\inst{1}}

   \institute{UMR 7293 J.L. Lagrange Laboratory, Universit\'{e} de Nice Sophia Antipolis, CNRS, Observatoire de la C\^{o}te d'Azur, Campus Valrose, 06108 Nice, France\\
         \and
         Astronomical observatory Belgrade, Volgina 7, 11060 Belgrade, Serbia\\
              \email{milic@aob.rs}
             }

   \date{}
   \titlerunning{Line polarization in rotating disks}
   \authorrunning{Mili\'{c} I. \& Faurobert M.}


  \abstract
   {To interpret observations of astrophysical disks it is essential to understand the formation process of the emitted light. If the disk is optically thick, scattering dominated and permeated by a Keplerian velocity field, Non-Local Thermodynamic Equilibrium  radiative transfer modeling must be done to compute the emergent spectrum from a given disk model.}
  {We investigate Non-local thermodynamic equilibrium polarized line formation in different simple disk models and aim to demonstrate the importance of both radiative transfer effects and scattering as well as the effects of velocity fields.}
   {We self-consistently solve the coupled equations of radiative transfer and statistical equilibrium for a two level atom model by means of Jacobi iteration. We use the short characteristics method of formal solution in two-dimensional axisymmetric media and compute scattering polarization, that is $Q/I$ and $U/I$ line profiles, using the reduced intensity formalism. We account for the presence of Keplerian velocity fields by casting the radiative transfer equation in the observer's frame form.}
   {Homogeneous, isothermal, self-emitting disks show complex intensity profiles which owe their shape to the interplay of multidimensional non-local thermodynamic equilibrium (NLTE) radiative transfer and the presence of rotation. The degree of scattering polarization is significantly influenced not only by the inclination of the disk with respect to observer, but also by the optical thickness of the disk and the presence of rotation. Stokes $U/I$ shows double-lobed profiles with amplitude which increases with the disk rotation.}	
   {Our results suggest that the line profiles, especially the polarized ones, emerging from gaseous disks differ significantly from the profiles predicted by simple approximations. Even in the case of \textbf{the} simple two-level atom model we obtain line profiles \textbf{which are diverse in shape, but typically symmetric in Stokes $Q$ and antisymmetric in Stokes $U$.} A clear indicator of disk rotation is the presence of Stokes $U$, which might prove to be a useful diagnostic tool. We also demonstrate that, for moderate rotational velocities, an approximate treatment can be used, where non-local thermodynamic equilibrium radiative transfer is done in the velocity field-free approximation and Doppler shift is applied in the process of spatial integration over the whole emitting surface.}


   \maketitle


\section{Introduction}

Proper interpretation of spectroscopic and spectropolarimetric observations of different kinds of gaseous disks relies on the understanding of the process of spectral line formation. Disks which scatter incoming radiation and/or emit by themselves are often encountered: circumstellar and circumbinary disks, protoplanetary disks, accrection disks on different scales, etc. It is not rare that understanding the role of the disk is crucial for a proper interpretation of the observations. The best known example is probably the unified model of active galactic nuclei as a result of detection of scattering polarization in spectral lines by \citet{AM85}. Another case where the presence of a disk proved essential is the explanation of the presence of emission lines in Be stars \citep[for a recent review of the subject, see:][]{Be_review_13} by the process of light scattering on the gaseous disk which surrounds the star.

Although modern radiative transfer techniques are almost routinely employed in solar and stellar physics \citep[for a review, see for example,\,][]{Carlsson_review_09}, detailed radiative transfer computations of line transfer in astrophysical disks are rarely found in the modern literature. \citet{Adam90} and \citet{Papkalla95} were among the first to employ numerical radiative transfer techniques in order to self-consistently solve the NLTE (non-local thermodynamic equilibrium) line formation problem in multi-dimensional disk models. A study of scattering line polarization in disks has been conducted by \citet{Ignace00}, but was limited to optically thin case, i.e. neglecting radiative transfer effects taking place in the disk. \citet{Vink05} used Monte Carlo simulations to investigate line scattering in rotating circumstellar disks. In their model, the disk is illuminated by a host star. They found that the presence of the ``gap'' between the surface of the star and the inner boundary of a rotating disk changes the way the degree of rotation of polarization plane varies with wavelength across the line profile. \textbf{Recently, \citet{WW12} confronted high-quality spectropolarimetric observation with detailed radiative transfer modeling \citep[see also][]{Carc2009, CB2008}.} Another recent study, conducted by \citet{Halonen13} dealt with the polarization of the continuum radiation in realistic disk models and demonstrated that it depends strongly both on the inclination of the disk as well as on the wavelength. \textbf{We also want to draw attention to the work of \citet{DH11} who deal with a different topic (polarization in the supernovae ejecta) but are relevant to this work as they also consider axial symmetry and the effects of scattering.}

Concerning the importance of detailed radiative transfer (RT) computations of spectral line formation, \citet{EAC12} have demonstrated the need for understanding the role of NLTE (i.e. line scattering) processes. Their main conclusion is that the double-peaked profiles, which are usually assumed to be an indicator of disk rotation, may occur simply due to the gradient of the source function (that is, the emission coefficient) along the line of sight. It is well known from the theory of stellar atmospheres that  scattering processes can cause these source function gradients, even in an isothermal medium. Another important consequence of line scattering is the so called scattering polarization. It is linear polarization, resulting from uneven population of Zeeman sub-levels. It is further influenced by  weak magnetic fields (Hanle effect). Scattering polarization and Hanle effect are routinely exploited in solar magnetic field diagnostics \citep[see, for example,\,][]{TB_Nature_04} where very complex and detailed formalism is utilized to compute the emergent Stokes vector. 

To authors' knowledge, the formalism of NLTE polarized line transfer \citep[the ultimate reference in the field is the monograph by][]{LL04}, has not yet been applied to polarized line formation in gaseous astrophysical disks. Here we attempt to make the first steps in such investigation. We self-consistently solve the radiative transfer and statistical equilibrium equations in the presence of line scattering in order to compute emergent polarized line profiles. We restrict ourselves to the case of a two-level atom line transfer, and to axisymmetric disk models. Our primary aim is not to obtain maximum realism but rather to investigate the importance of radiative transfer effects and disk rotation on emergent line profiles. We thus, in this paper, restrict ourselves to idealized, homogeneous \textbf{(in the context of this work meaning: constant density/opacity)}, isothermal disk models. One is to keep in mind that even in these very simple cases, we are faced with the interplay of radiative transfer effects in optically thick media, non-LTE effects, multidimensionality effects and the influence of the disk differential rotation.  All of these effects, as we shall see in the remainder of the paper, leave a significant imprint on the emergent line shape and line polarization.

Firstly, in Section\,2 we outline the NLTE problem of the second kind and the method of solution. In sections\,3 and 4 we present results for self-emitting and illuminated disks, respectively. In Section\,5 we draw some conclusions and discuss on possible directions for future work.

\section{Method}

We are interested in obtaining the emergent polarized intensity from the disk as a whole. What we observe is spatially (and, in case of photometry, frequency-) integrated intensity, or, in general, Stokes vector. To compute the total disk radiation, one has to compute the specific emergent intensity and to perform spatial integration over the whole emitting surface.  

The majority of the recent works dealing with disk-like objects utilizes Monte Carlo methods to compute disk images and/or SED (spectral energy distribution) curves. However, they are mostly aimed at computing the continuum radiation as well as self-consistent disk temperature distribution. We are interested in a different process, \emph{line scattering}, so we consider the disk model (opacity and temperature distribution) to be given, that is, we assume that spectral lines do not influence significantly the energy balance of the object. We compute the emergent intensity by solving the radiative transfer equation (RTE), which in the most general case reads:
\begin{align}
\frac{\partial \hat{I}(\vec{r}, \vec{\Omega}, \nu)}{\partial t} + \nabla \hat{I}(\vec{r}, \vec{\Omega}, \nu) = & \nonumber \\
= \hat{\eta} (\vec{r}, \vec{\Omega}, \nu) - \hat{\chi} (\vec{r}, \vec{\Omega}, \nu) \hat{I}(\vec{r}, \vec{\Omega}, \nu). &
\label{RTE_0}
\end{align}
Here $\hat{I}$ is the polarized intensity given as the Stokes vector $(I,Q,U,V)^{\dagger}$, $\hat{\eta}$ is the emission coefficient (also a four-vector) and $\hat{\chi}$ is the $4\times4$ absorption matrix. $\vec{\Omega} = (\theta, \varphi)$ is the direction of the propagation and $\nu$ is the  radiation frequency. 

\textbf{We assume axial symmetry and thus cast the radiative transfer equation in a 2D cylindrical coordinate system $(z,r)$. In the case where we also introduce a Cartesian coordinate system, the $z$ axis coincides with axis of the cylinder while the disk lies in the $xy$ plane. In this work, we are interested in obtaining the emergent polarized intensity at the top boundary of the disk ($z=z_{total}$), as we restrict ourselves to  disks which are geometrically thin (i.e. $r_{total} \gg z_{total}$, where subscript ``total'' refers to the size of the disk along the appropriate coordinate) so the emitting surface is actually the top boundary.}

\textbf{To obtain the emergent (generally, polarized) intensity, we need to solve the radiative transfer equation.} $\hat{\eta}$ and $\hat{\chi}$ depend on the populations of the atomic energy levels, which in turn depend on temperature and density, but also on the radiation field itself, which brings inevitable coupling in the radiative transfer problem. We will present the way of dealing with this problem under the following assumptions:
\begin{itemize}
\item We restrict ourselves to time-independent (stationary) problem and to axisymmetric objects. We thus cast the radiative transfer equation in 2D cylindrical coordinate system. \textbf{The   spatial dependence of all the quantities is given by $z$ and $r$ coordinates. $z$ goes from zero to $z_{\rm{total}}$ where $z_{\rm{total}}$ corresponds to the top boundary of the disk. $r$ goes from zero to $r_{\rm{total}}$ where $r=0$ refers to the center of the disk.}
\item We assume that line transport is taking place in a two-level atom model where the lower level of the transition is \emph{unpolarized}. Under this assumption the absorption matrix takes the form $\hat{\chi} = \chi \vec{1}$, where $\vec{1}$ is unit matrix. The  two-level atom assumption is adequate for  resonance lines. Polarized line formation process for transitions requiring multi-level treatment (such \textbf{as} H$\alpha$) is much more complex but the main effects outlined in this work should manifest themselves similarly.
\item We consider a magnetic field - free case. That means that there is no Hanle effect \textbf{\citep[see][for optically thin computations of Hanle effect in disks]{Ignace_Hanle}} and that Stokes $Q$ and $U$ arise only due to the anisotropy of the radiation. Using the formalism from \citet{AnushaIII}, we can easily implement the Hanle effect due to  large scale magnetic fields in our computations (but note that the magnetic field would have to be axisymmetric). Also, we neglect any Zeeman effect, i.e.  Stokes $V$ is identical to zero. 
 \end{itemize}
With these assumptions, $\chi$ does not depend on the direction any more. We divide Eq.\,\ref{RTE_0} with $-\chi$ and use the so called along-the-ray form of the radiative transfer equation to get:
\begin{equation}
\frac{d \hat{I}(z,r, \vec{\Omega}, \nu)}{d \tau} = \phi(\nu) \left [ \hat{I} (z,r,\vec{\Omega}, \nu) - \hat{S}(z,r,\vec{\Omega}, \nu)\right ] .
\label{RTE}
\end{equation}
Here, $\hat{S}$ is known as the \emph{source function}, the ratio between emission and absorption coefficient and $\phi(\nu)$ is the line absorption profile \textbf{which describes the dependance of the line opacity  on frequency}. $\tau$ is the line-integrated optical depth along the ray, that is:
\begin{equation}
d \tau = -\chi_{\rm{l}} ds,
\end{equation}
Where $\chi_{\rm{l}}$ is the line-integrated absorption coefficient and $ds$ is the elementary path along the ray. Here the  ``ray'' is a line with the orientation given by the angle $\vec{\Omega}$, along which the radiative transfer equation is integrated. Under the assumption of complete frequency redistribution in the line scattering process (frequencies of incoming and outgoing photons are uncorrelated), the source function becomes independent of frequency and has the following fom:
\begin{align}
\hat{S}(z,r, \vec{\Omega}) = \epsilon \hat{B}(r,z) + (1-\epsilon)\hat{W} \times \nonumber \\
\int_{-\infty}^{\infty} \phi(\nu) d\nu \oint \frac{d \vec{\Omega}' }{4\pi} \hat{P}(\vec{\Omega}, \vec{\Omega}') \hat{I}(z,r,\vec{\Omega}', \nu),
\label{SE}
\end{align}
where $\epsilon$ is the photon destruction probability, $\epsilon = C_{ul}/(A_{ul} + C_{ul})$ , $A_{ul}$ and $C_{ul}$ being the upper level radiative and collisional lifetime, respectively, and $\hat{B}=(B,0,0)^{\dagger}$ where $B$ is equal to the local Planck function at the frequency $\nu$, $\hat{W}$ is a diagonal matrix whose elements account for the intrinsic line polarizability and depolarization due to the presence of elastic collisions and/or unresolved, randomly oriented magnetic field. \textbf{$\hat{P}(\vec{\Omega}, \vec{\Omega}')$ is known as the scattering matrix which describes correlations between incoming and outgoing photon.}  Equation \ref{SE} is known as the \emph{equation of  statistical equilibrium} for a two level atom.

Smaller the rate of collisions ($C_{ul}$) with  respect to the radiative rate ($A_{ul}$), greater the  importance of the scattering processes and the source function differs more from the local Planck function. For high rate of collisions, the source function is practicaly identical to the local Planck function, and this situation is known as local thermodynamic equilibrium or LTE. The importance of departures from LTE (NLTE) is a well known problem in the field of stellar atmospheres \citep{Mihalasbook}. The non-LTE case, generalized to the problem of line scattering polarization is known as the ``NLTE problem of the second kind'' \citep[for a recent review, see, for example][]{TB09}. 

We need a self-consistent solution of coupled Eqs.\,\ref{RTE} and \ref{SE} on a 2D ($z,r$) grid, under the assumption of axial symmetry, and some assumed angular and frequency discretization. A detailed description of a numerical solution in this kind of geometry is given in \citet{Milic13}, where the NLTE problem of the second kind is cast in the so called real reduced intensity formalism first introduced by \citet{Frisch07} and generalized to 3D geometry by \citet{AnushaI}. In the following subsection we briefly recap the reduced intensity formalism and refer the interested reader to works cited above.

\subsection{Reduced intensity formalism}

It can be shown \citep{Frisch07, AnushaI}, that both the polarized source function $\hat{S}$ and the polarized intensity $\hat{I}$ can be decomposed by using the so called real reduced source function and reduced intensity \textbf{($\hat{\cal S}$ and $\hat{\cal I}$, respectively)}:
\begin{equation}
\hat{S} = \hat{T}_S \hat{\cal S}
\end{equation}
and 
\begin{equation}
\hat{I} = \hat{T}_I \hat{\cal I}
\end{equation}
where ${\cal S}$ and ${\cal I}$ are the real reduced source function and intensity and the matrices $\hat{T}_S$ and $\hat{T}_I$  are linear operators that relate the reduced formalism and the Stokes formalism. The reduced formalism results from the expansion of the scattering phase matrix on the irreducible spherical tensors ${\cal T}_K^Q$ basis \citep[see][for details]{LL04}. The explicit form is given by \citet{Frisch07} for 1D media and by \citet{AnushaI} for the general, 3D case. The main advantage of this approach is that the reduced source function is now \emph{angle independent}, which significantly increases the efficiency of numerical codes used for the solution of the NLTE problem of the second kind. The reduced intensity and reduced source function obey the same form of radiative transfer equation:
\begin{equation}
\frac{d \hat{\cal I}(z,r, \vec{\Omega}, \nu)}{d \tau} = \phi(\nu) \left [ \hat{I} (z,r,\vec{\Omega}, \nu) - \hat{\cal S}(z,r)\right ],
\label{rrRTE}
\end{equation} 
while the statistical equilibrium equation has the following form:
\begin{align}
\hat{\cal S}(z,r) = \epsilon \hat{\cal B}(r,z) + (1-\epsilon)\hat{W} \times \nonumber \\
\int_{-\infty}^{\infty} \phi(\nu) d\nu \oint \frac{d \vec{\Omega}' }{4\pi} \hat{\Psi}(\vec{\Omega}') \hat{\cal I}(z,r,\vec{\Omega}', \nu),
\label{rrSE}
\end{align}
\textbf{where $\Psi(\vec{\Omega}')$ is the scattering matrix in the real reduced formalism.} The problem of coupled equations \ref{rrRTE} and \ref{rrSE} is formally identical to the standard (i.e. unpolarized) NLTE radiative transfer problem, and, with appropriate changes one can use the same numerical scheme to solve it. Once the self-consistent solution has been found, one can compute the emergent Stokes components of interest from the real reduced intensities \citep[see][for the exact form of the transformation]{AnushaI}.

We solve the problem of the coupled equations of radiative transfer and statistical equilibrium by using Jacobi iteration. This method is based on casting  Eq.\,\ref{RTE} (or, in the polarized case, Eq.\,\ref{rrRTE}) in an integral form, and introducing the so called $\Lambda$ operator so that:
\begin{equation}
\hat{\cal S} = \epsilon \hat{B} + (1-\epsilon) \hat{\Lambda} [\hat{\cal S}].
\end{equation}
In Jacobi iteration, $\hat{\Lambda}$ is split into a local (diagonal) and a non-local part. The local part is easily inverted and the errors introduced by this approximation are iteratively corrected. For a detailed review of the subject, see \citet{Hubeny03} and for applications of Jacobi iteration and similar techniques to polarized NLTE radiative transfer see \citet{TB03}. The details on applying Jacobi iteration to the polarized line formation in reduced intensity formalism for this specific geometry are given in \citet{Milic13}. 

For the research conducted here, we have slightly changed the formal solution (way of computing the specific intensity from a given value of the source function) scheme used in \citet{Milic13} so it now corresponds to BESSER \citep{PORTA13} formal solution which utilizes second order, smooth and monotonic interpolation technique. We have employed BESSER interpolation also for spatial and angular interpolation needed in order to perform the formal solution along-the-ray. We also use the sub-gridding strategy proposed by \citet{VNHL02} and subsequently by \citet{IRIS13}. This enables more precise computation of the monochromatic optical path along the ray in the observer reference frame (see next subsection).

\subsection{Velocity fields}

To compute the emergent spectrum \textbf{for} rotating disks, one must properly account for the rotation of the disk. In the case of rigid body rotation, any two points on the disk do not move with respect to each other. That means that the radiative transfer can be done in the static case and radiation from different parts of the disk can be added together afterwards (see subsection 3.1). A more realistic, and, from the radiative transfer point of view, more problematic case is when the gas in the disk is following a Keplerian velocity distribution:
\begin{equation}
v(r) = \sqrt{\frac{GM}{r}}.
\end{equation}
Or, if we define the rotational velocity $v_0$ at radius $r_0$, then:
\begin{equation}
v(r) = v_0 \sqrt{\frac{r_0}{r}}.
\label{rot}
\end{equation}
Now the situation is clearly different from the rigid body rotation as any two points in the disk are moving with respect to each other. Since the radiative transfer problem is a spatially coupled problem (i.e. the radiation field in one point depends on the radiation field in other points), Doppler shifts resulting from this relative movement must be taken into account. Here, we have used the observer's frame formalism \citep[see, e.g.\,][]{Mihalasbook}, where all the relevant quantities are expressed in the coordinate frame of the observer. All the equations keep the same form, but frequencies and directions become coupled. For example, the opacity coefficient (and, subsequently, the optical path distance between two points) is now also direction-dependent, as:
\begin{equation}
\chi = \chi_0 \phi(\nu'), 
\label{obs_frame}
\end{equation} 
where 
\begin{equation}
\nu' = \nu \left(1-\frac{\hat{\vec{s}} \cdot \vec{v}}{c}\right).
\end{equation} 
Here $\vec{s}=(\sin \theta \cos \varphi, \sin \theta \sin \varphi, \cos \theta)$ and $\vec{v} = (v_x, v_y, v_z)$, while the observer is assumed to be situated in $(0,0,0)$. This formalism is straightforward and powerful, as it allows for inclusion of arbitrary velocity fields but suffers from computational inefficiency when large velocities come into play as it requires the use of a frequency mesh which covers all possible doppler shifts. For rapidly rotating disks, this might easily render the computation rather cumbersome. However, it turns out that the relevant quantity is not the velocity itself but rather the ratio of the velocity to the Doppler velocity of the gas.  To clarify this, let us use the reduced frequency scale instead of regular frequency, i.e. substitute $\nu$ with $f$:
\begin{equation}
f = (\nu - \nu_0) / \Delta\nu_D,
\end{equation}
where $\Delta\nu_D = \nu_0 v_D / c$ is known as the Doppler width of the line. $v_D$ is then the Doppler velocity which accounts for all the random, small-scale velocities of the gas (thermal velocities of constituting atoms and small scale turbulent velocities). Eq.\,\ref{obs_frame} now takes the form:
\begin{equation}
\chi = \chi_0 / \Delta v_D \,\phi (f - \hat{\vec{s}} \cdot \tilde {\vec{v}}),
\end{equation}
where $\tilde{\vec{v}}$ is the velocity given in units of the Doppler velocity. It is clear now that the relative Doppler shift is what is important. Therefore, large random velocities can suppress the effect of rotation and, in principle, of any systematic velocity field. This statement is an extremely important thing to keep in mind. Here we restrict ourselves to the case of moderate rotational velocities ($v_{\rm{rot}} \approx 10 v_D$). This is because we are primary interested in the study of circumstellar disks, but the results and our main points should be applicable to faster rotating disks as well. This regime of intermediate velocities is particularly interesting because velocities are large enough to influence the line formation process\footnote{In the cases when macroscopic velocities are smaller than the Doppler velocity of the gas, they have no effect on the line formation and hence cannot be detected.}, but still not large enough to justify, e.g. the Sobolev approximation or similar high-velocity regime. 

To have some idea of the velocities we consider here, one should keep in mind that the thermal velocity of hydrogen atoms on temperature of $\approx 10^4$\,K is around $10$\,km/s. Turbulent gas motions could increase this velocity somewhat, that means that the rotational velocities considered in the next chapter are of the order of $100$km/s. If we consider a typical UV or optical line with $\lambda=300-800$\,nm, one Doppler width would correspond to $\approx 0.01$\,nm, which requires a spectral resolution of few times $10^4$. 

 \section{Results}
 
\textbf{In this section we show the results of applying the method described above to two models of gaseous disks. Both are homogeneous and isothermal. First we consider disks which are completely self-emitting, i.e. the light of the central body is neglected. The other type are disks which have a star in the center. In our computations we properly incorporate the starlight in the boundary conditions for the radiative transfer problem but we \emph{do not add} the light from the part of the star which is not occulted by the disk to the total flux. In the remainder of the chapter we further comment on both of the models.} 
 
Before inspecting the profile shapes for different disk models it is important to realize that the velocity field influences the emergent spectrum in two ways: i) Firstly, velocity gradient, if large enough, can alter the distribution of the source function in the object, as the photons can escape the medium more easily; and ii) The gradient of the velocity field along a line of sight causes Doppler shifts, thus changing the effective depth until which the medium is probed \footnote{The rule thumb in radiative transfer is that the emergent radiation at frequency $\nu$ is equal to the source function at monochromatic optical depth $\tau_{\nu}$. The presence of velocity fields allows one to ``probe'' different regions than in the static case. Because of the spatial variations of the source function this generally leads to different emergent profiles.}. For moderate velocities like this we expect the second effect to be dominant. 

However, we first investigate the decrease of the source function due to the loss of photons because of the velocity gradient. We consider a disk model with its radial line-integrated optical thickness $\tau_r$ equal to 1000 and vertical optical thickness $\tau_z$ equal to 10. We set $\chi_0 = 1$ and thus the geometrical grid is identical to the optical depth grid. The Planck function is set to unity (i.e. the disk is isothermal), and the photon destruction probability is equal to $\epsilon = 10^{-4}$ everywhere in the disk. In this and all other computations we assume a Doppler line absorption profile, i.e. $\phi(x) = \exp(-x^2)/\sqrt{\pi}$. \textbf{We do not account for continuum opacity, i.e. we are interested only in the line radiative transfer.} \textbf{Also, as the opacity is constant, we use the optical depth as the relevant "spatial" coordinate. In the remainder of the text, the optical depth/thickness of the disks is computed from the so called mean or line integrated opacity, $\chi_0$.}

The disk is rotating according to the following law:
\begin{equation}
v(r) = v(r_{\rm{total}}) \sqrt{\frac{r_{\rm{total}}}{r}},
\end{equation}
for $r > r_c$, where $r_c = r_{\rm{total}}/10$ and
\begin{equation}
v(r) = v(r_c) \frac{r}{r_c}
\end{equation}
for $r<r_c$. That is, the disk is rotating like a solid body in the inner $10\%$ part of the radius (this is to avoid very large velocities for small values of the radius) and with a Keplerian law elsewhere. \textbf{For now, we assume that the disk is emitting by itself only, i.e. the radiation of the central object is assumed to be negligible.}

\begin{figure}
\centering
\includegraphics[height = 0.49\textwidth, angle = 90]{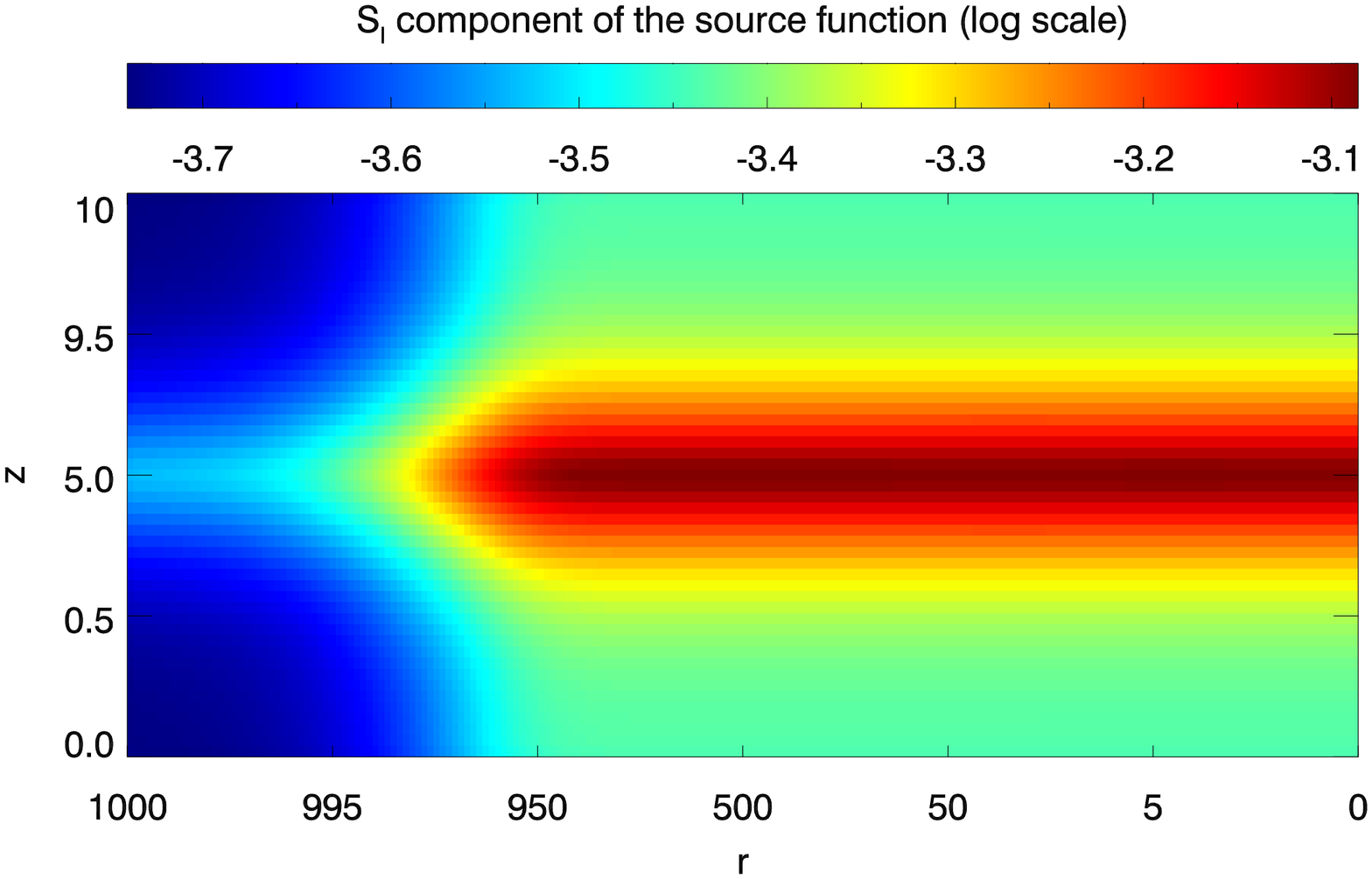}
\includegraphics[height = 0.49\textwidth, angle = 90]{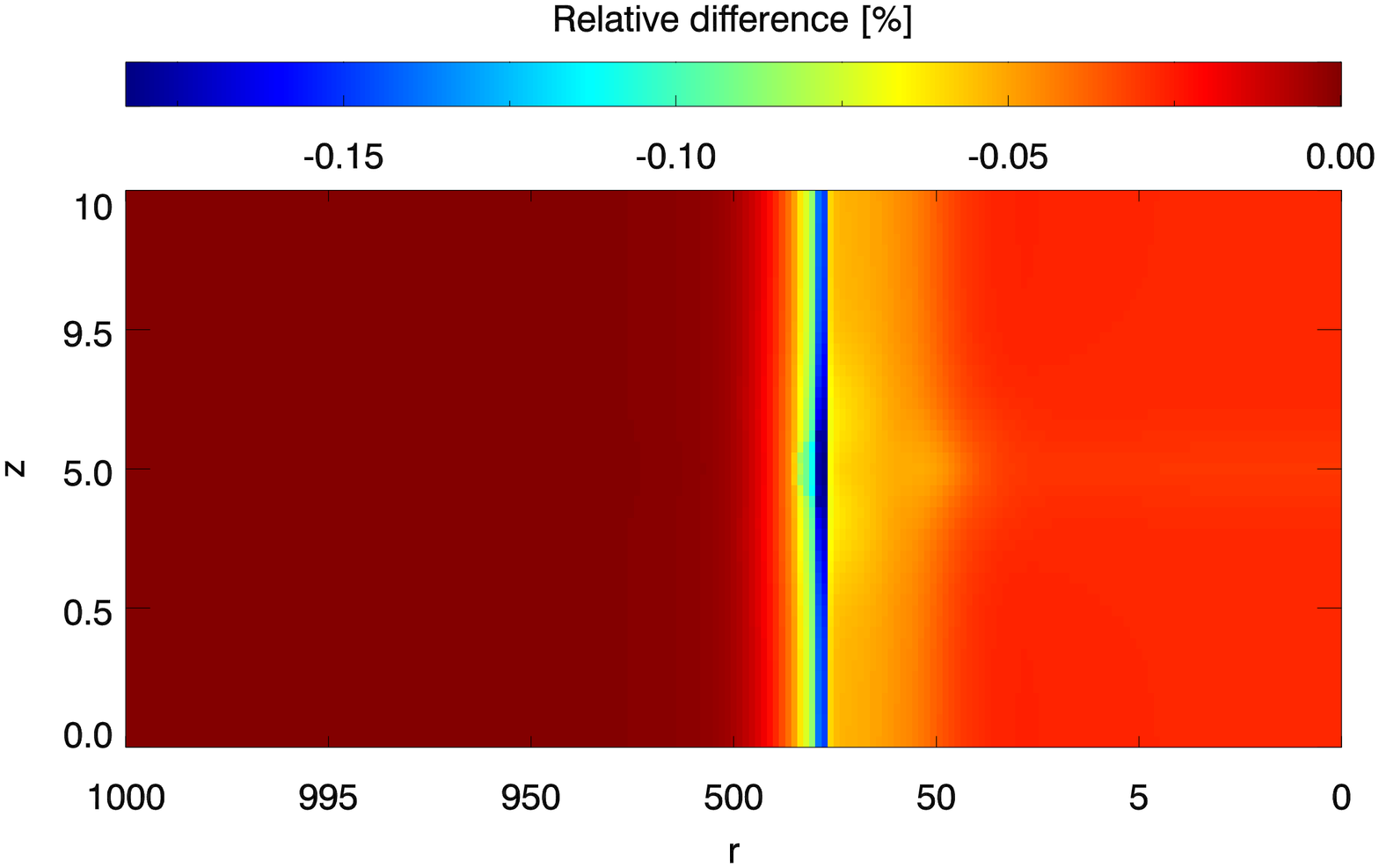}
\caption{2D distribution of $S_I$ source function in the $z,r$ plane for the non-rotating cylinder (up) and relative difference between rotating and static case (down). The source function is given in units of the Planck function. Note that both $z$ and $r$ axis are double-logarithmic, i.e. the sampling is finer toward the disk boundaries.}
\label{s_distribution_1}
\end{figure}

We self-consistently compute the polarized source function both in the static case and the case when $v(r_{\rm{total}}) = 5 v_D$ and compare the values of  ${\cal S}_0$ (first component component of the reduced source function, which is to a good approximation identical to the first component of the Stokes source function, $S_I$) between the two cases.  The 2D distribution of the source function in both cases is shown in Fig.\,\ref{s_distribution_1}. Here the total optical thickness along the $z$ and $r$ coordinates is equal to $10$ and $10^3$ respectively. Note that, although the optical thickness is substantial and far from the optically thin case, the medium is \emph{effectively thin}, i.e. the optical thickness of the medium along both axis is much smaller than the thermalization length for  the line source function ($S_I$ would be equal to the Planck function at optical depths on the order of $ 1/\epsilon$). 

Detailed inspection reveals that the relative differences between the source function values in the non-rotating and rotating cases are smaller than $5\%$ in the greatest portion of the disk. We therefore attribute the differences between the computed spectra mostly to the Doppler shift of the radiation due to the rotation of the cylinder. Note that the emergent profiles can have non-trivial shapes even in the static case, as the NLTE effects come into play \citep[see the detailed discussion in][]{EAC12}. As is evident from Fig.\,\ref{s_distribution_1}, the source function varies both with depth and with the radius. Thus, we are to expect complicated profile shapes, possibly with multiple peaks, as ones encountered in \citet{EAC12}. We want to stress that the major difference between this work and theirs is in the fact that we completely self-consistently take into account multidimensional radiative transfer effects, and consider line polarization as well. We now discuss on the shape of emergent spectral lines for different rotational velocities. 

\begin{figure}
\centering
\includegraphics[height = 0.49\textwidth, angle = 90]{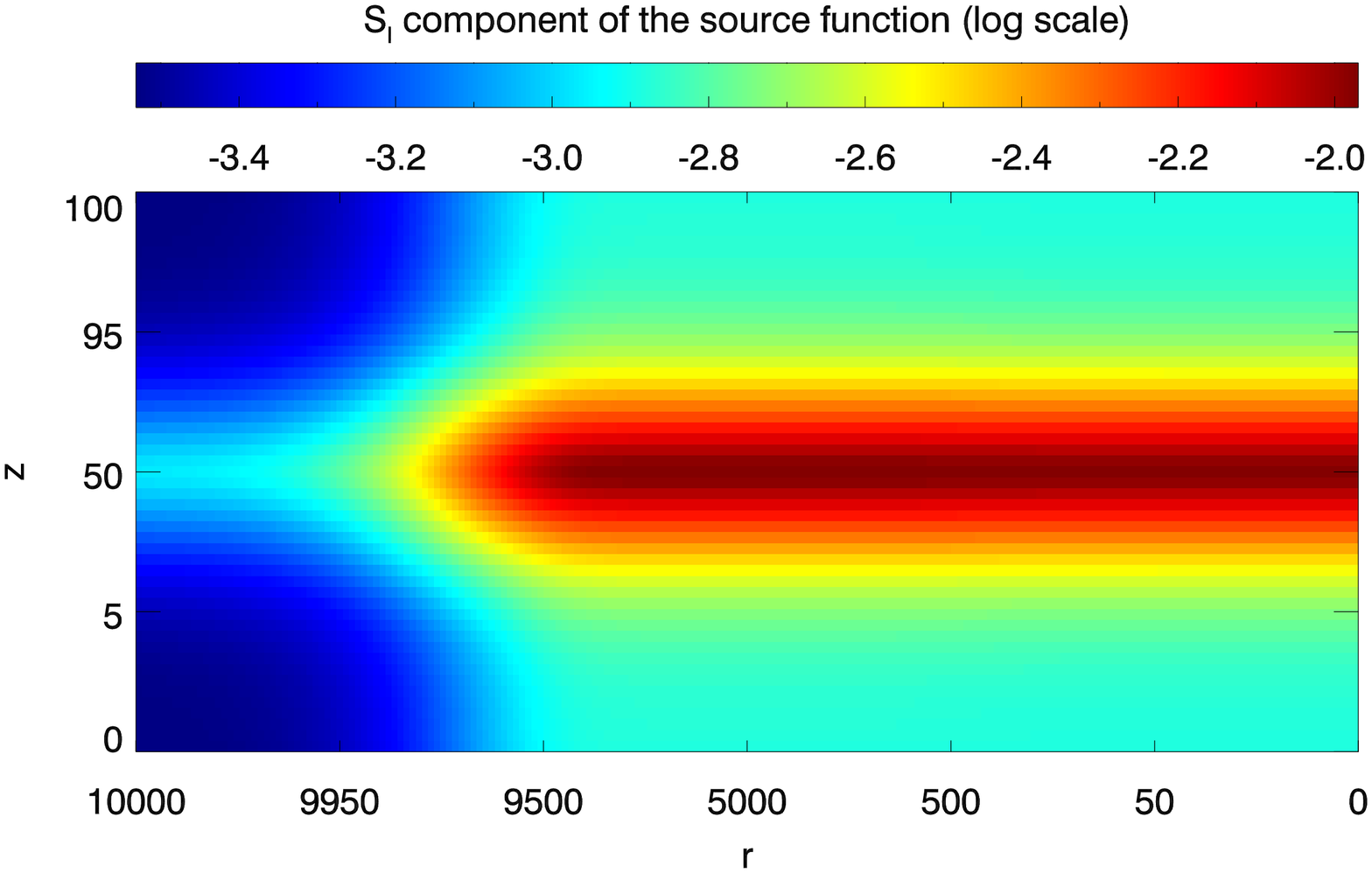}
\includegraphics[height = 0.49\textwidth, angle = 90]{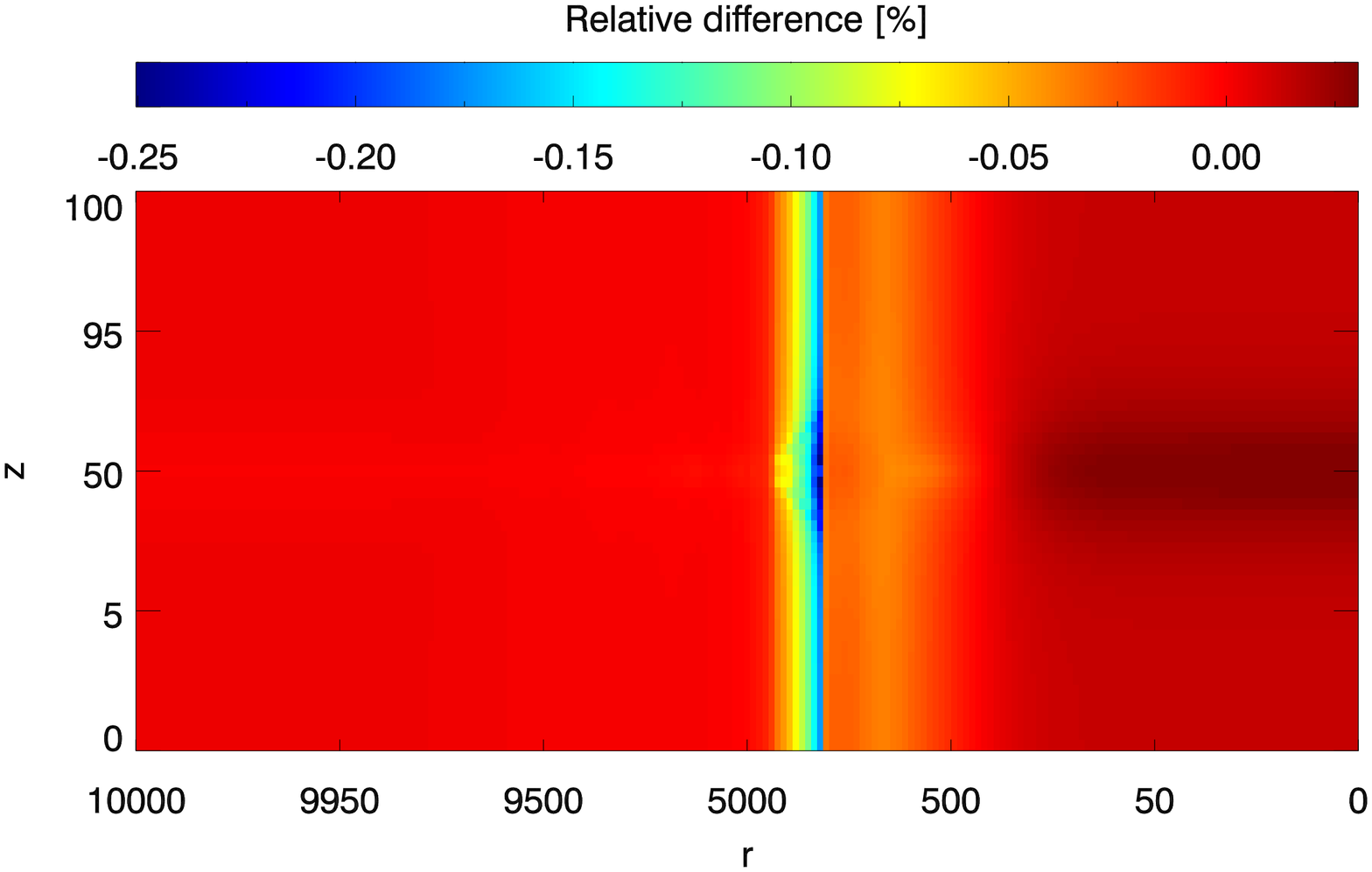}
\caption{Same as in Fig.\,\ref{s_distribution_1}, but for a ten times more opaque cylinder.}
\label{s_distribution_2}
\end{figure}

Fig.\,\ref{s_distribution_2} shows the spatial distribution of the ${\cal S}$ source function first component for 10 times more opaque cylinder (\textbf{i.e. opacity is equal to ten everywhere in the disk}, $\tau_r = 10^4$ and $\tau_z = 10^2$). \textbf{We now witness a larger span (almost two orders of magnitude) of the source function inside the medium, and, correspondingly, different shapes of line profiles are expected.} As different wavelengths in the spectral line probe different optical depths, we are to expect significant difference in line center and line wings, at least in the static case. It is very important to understand all the effects of non-constant source function (which is here only due to scattering), as the shape of the emergent line is uniquely determined by the distribution of the source function along the line of sight and the emission profile of the spectral line \citep[we again point the reader to][for an in-depth discussion]{EAC12}. \textbf{An interesting feature in both figures is a thin vertical stripe close to the half of the disk radius. We believe that this is a numerical artifact due to the use of the logarithmic grid for the spatial discretization. It then creates a relatively large jump in rotational velocity between two consecutive points and an "artificial" loss of photons.}

Once the (polarized) source function is obtained, one can perform a single formal solution for given directions and frequencies in order to compute the shape of the emergent spectral line.

\subsection{Formal solution of the radiative transfer equations in a rotating disk}

Due to the insensitivity of the source function distribution to this, relatively slow rotation, we can conclude that one can use the source function distribution from the static case and just perform a formal solution in desired directions and on desired frequencies, for different velocity fields. However, this is where awkwardness of a cylindrical coordinate system comes into play. Let us first notice that, if we want to obtain the spatially integrated Stokes vector from the top of the cylinder (cylinders we consider here are geometrically thin, i.e. $r \gg z$), one has to integrate the emergent Stokes vector over all radii and azimuths:
\begin{equation}
\hat{I}^{\rm{emergent}}(\theta, \nu) = \int_0^{r_{\rm{total}}}  r\,dr \int_0^{2\pi} d\varphi \hat{I}(z_{\rm{total}}, r, \theta, \varphi, \nu).
\label{integration}
\end{equation}
We recall that, in the case of Cartesian coordinates, one performs a similar integration along $x$ and $y$. It can be seen that the \emph{angular} coordinate $\varphi$ which describes the direction of radiation plays the role of the spatial coordinate which describes the azimuth of the specific point on the disk. This is, of course, due to the axial symmetry of the medium. In practice, one needs many discrete azimuths to compute the emergent intensity in order to properly sample all line-of-sight velocities, which decreases the efficiency and time saving brought by treating this problem in a cylindrical 2D coordinate system \citep[for discussion on advantages and drawbacks of numerical radiative transfer on 2D cylindrical meshes see][]{Milic13}.

We can, however, notice that in this case the radiative transfer effects occur mostly along $z$ and $r$. This means that one can compute the emergent spectrum in a rotating case by computing the emergent Stokes intensity in the static case and then performing the integration (Eq.\,\ref{integration}) over $r$ and $\varphi$ while properly red/blue-shifting the contribution of each point. That is:
\begin{align}
\hat{I}^{\rm{emergent}}(\theta, \nu) = \int_0^{r_{\rm{total}}} r\,dr \int_0^{2\pi} d\varphi  \times \nonumber \\
\hat{I}^{\rm{static}}(z_{\rm{total}}, r, \theta, \varphi, \nu - v(r)\cos\varphi).
\label{integration_app}
\end{align}

\textbf{This method is a mixture of exact (computation of the source function) and approximate (formal solution) approach \citep[for another method which combines two different approaches see][]{Lamers1987}.} We have compared the results obtained this way with the ones computed by performing the formal solution in the rotating case using the observer's frame formalism and we have found no significant differences (See Appendix A for the illustration.). This means that: a) Equation\,\ref{integration_app} can be used to compute the Stokes profiles emergent from the disk, which saves time and makes the investigation of different rotation laws easier, because one only needs to perform one self-consistent NLTE solution and one formal solution in the static case while the inclusion of different rotation laws \textbf{becomes a matter} of post-processing; b) Our assumption that lateral radiative transfer effects are negligible really does hold for this type of disks which means that we could even attempt to emulate non-axisymmetric disks by "stitching together" pieces of different axisymmetric disks for which we have computed the emergent intensity separately. \textbf{Such a computation, however, would require a detailed prior investigation as the dependance of the physical parameters (temperature, density) on the azimuth could lead to increase the importance of lateral radiation transport.}

We now inspect the emergent spectral line profiles for very simple disk models, mostly to demonstrate the effects of rotation on emergent scattering polarization profiles. All our models are isothermal, have constant opacity equal to unity and constant photon destruction probability equal to $\epsilon = 10^{-4}$. This is a standard value of $\epsilon$ used in academic test problems in NLTE radiative transfer computations in stellar atmosphere modeling. Constant opacity implies that the geometrical scale is identical to the optical path scale. Low value of $\epsilon$ implies that the source function is dominated by the mean intensity of the radiation field, rather than by the Planck function.

We consider four different disk models in total: they come in two different opacities and we also inspect disks which are only self-emitting as well as ones which are illuminated from the inside by a host star. \textbf{The only source of radiation in the self-emitting case is the thermal energy of the gas in the disk, while in the case of illuminated disks, the main source is actually the starlight which is scattered in the disk.} The radius of the star is equal to $\frac{1}{5}$ of the disk radius and it emits like a blackbody \textbf{(the star temperature is the same as the disk temperature)}. Note that, although the star is emitting isotropically, the disk is still illuminated \emph{anisotropically}. For readers' convenience we name these four models: $A$, $B$, $A_i$, $B_i$. We consider all four models in the static and rotating cases. The properties of the models are summed in Table \ref{t1}. \textbf{In the subsequent results we normalize the emergent Stokes I with respect to the area of the top surface times the Planck function, i.e. with respect to $R_{\rm{total}}^2 \pi B$.}

\begin{table}
\begin{center}
\caption{Disk models used to compute emergent Stokes spectra. All models are isothermal, with opacity equal to unity, and $\epsilon = 10^{-4}$.}
\begin{tabular}{c  c  c  l }
  Model & $\tau_r$ & $\tau_z$ & Illumination  \\
  \hline 
  A & $10^3$ & $10$ & Self-emitting \\
  B & $10^4$ & $10^2$ & Self-emitting \\
  A$_i$ & $10^3$ & $10$ & Self-emitting + Star illumination \\
  B$_i$ & $10^4$ & $10^2$ & Self-emitting + Star illumination \\
\end{tabular}
\end{center}
\label{t1}
\end{table}
 
 \subsection{Self-emitting disks}

\begin{figure*}
\centering
\includegraphics[width = 17cm]{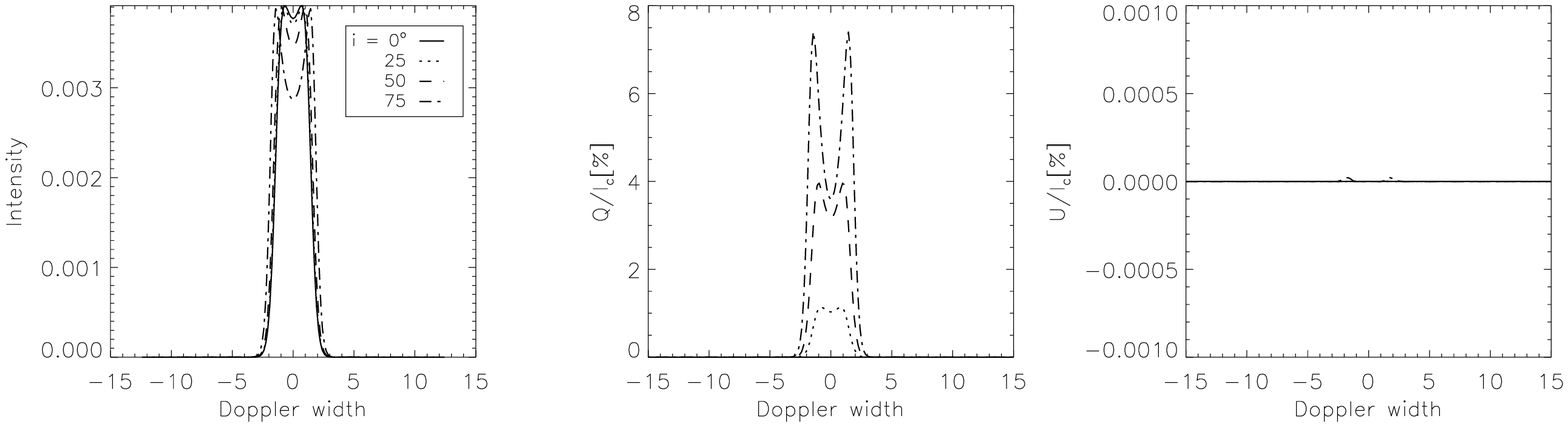}\\
\includegraphics[width = 17cm]{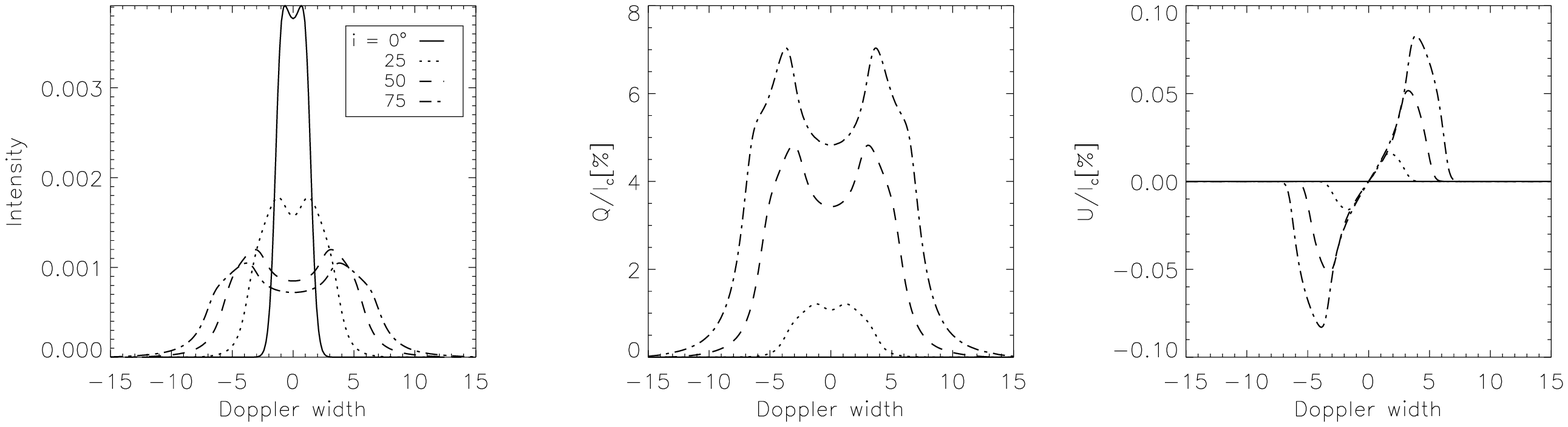}
\caption{Polarized line profiles for self-emitting cylinder of radial mean optical thickness equal to 1000 and vertical mean optical thickness equal to 10 (model A), for static (up) and rotating (down) case. Note that the rotation induces a broadening of the intensity and $Q/I_c$ profiles, as well as non-zero Stokes $U/I_c$. Intensity profiles are normalized to the total emitting surface.}
\label{profiles_model1_static}
\end{figure*}

Fig.\,\ref{profiles_model1_static} (top panel) shows the emergent profiles for several disk inclinations (inclination of $0^{\circ}$ indicates that the disk is viewed face-on) for the non-rotating disk. Profiles for different inclinations have similar shapes. The line center intensity decreases with increasing inclination, because we are probing  regions closer and closer to the surface, where the source function is smaller. This effect is similar to limb darkening effect in stellar/solar physics. As expected, $Q/I_c$ (we normalize the linear polarization with respect to stokes $I_c$ at  line center)  increases with increasing inclination (which, again, is expected, as it is well-known from solar observations that scattering polarization rapidly increases as we observe closer and closer to the limb). Profiles again have a central ``dip'' due to the interplay between spatial variations of the $S_Q$ and $S_I$ source function components. We expect Stokes $U$ to be zero, as the object is completely axisymmetric. Indeed, in our computations we find  that Stokes $U/I_c$ which is smaller than $10^{-5}$, and its (small) presence is due to slight numerical inaccuracies in the code. 

From the two bottom panels of Fig.\ref{profiles_model1_static} it is evident that Keplerian rotation leaves a significant imprint on the emergent profiles. Firstly, profiles viewed \textbf{face} on remain the same, which is expected as the values of  the source functions are changed negligibly, and for this viewing direction the Doppler effect has no influence as the rotation is taking place in the $xy$ plane. With increasing inclination, the intensity profiles become broader and eventually start exhibiting double peaks and non-Gaussian shapes which are the result of the joint action of Doppler shift and inhomogeneities \textbf{along $z$ and $r$} in the the source function $S_I$ because of NLTE effects. The usual impression that disks that rotate obeying the Kepler law also exhibit a double-peaked emission is not completely true for these moderate velocities. For $v_0 = 10 v_D$ and high inclinations we see some indication of double peaked profile, which still has a rather interesting spectral structure.

The situation is similar when we analyze $Q/I_c$ line profiles. They become broader, and exhibit double peaks, but, interestingly, they have approximately the same polarization in the center of the line, regardless of the rotational velocity. Their shapes are not easy to interpret as they are results of both NLTE effect, multidimensional radiative transfer effects and the presence of velocity fields. The slight ``splitting'' of the line can again be seen for high inclinations for the rotating case. The most interesting feature is the presence of non-zero Stokes $U$. It is an expected effect, already found in the paper of \citet{Vink05} and also explained and  by \citet{Smith05} in the context of spectropolarimetric studies of active galactic nuclei. Here we discuss further on the topic and explain the origin of this polarization signature.

Firstly it is important to understand that, in the absence of  large-scale magnetic fields, $Q/I$ and $U/I$ polarization probe the polar and azimuthal anisotropy of the radiation, respectively. In the absence of the \textbf{local} limb darkening/brightening \textbf{(i.e. if the radiation field does not vary with $\theta$)} Stokes $Q$ would be zero. The same stands for  Stokes $U$ in the absence of axial asymmetry. Now, note that the radiation field is, strictly speaking non-axisymmetric for any $r\neq0$. \textbf{At each point in the disk, the emergent stokes $U$ depends on (`probes') the amount of azimuthal anisotropy. }

\begin{figure}
\includegraphics[width = 0.49 \textwidth]{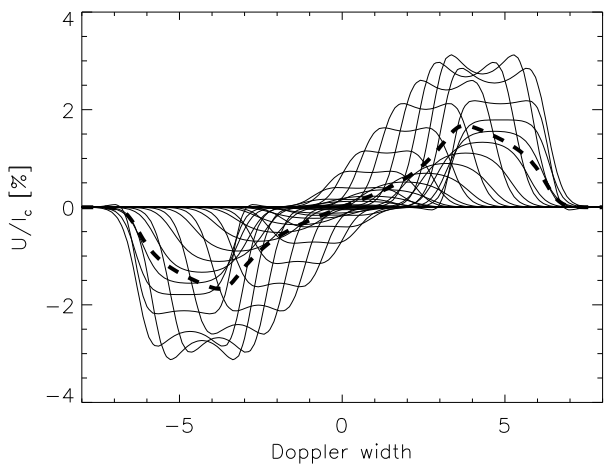}
\caption{Stokes $U/I_c$ profiles plotted for a fixed value of $r$ and different azimuths (thin solid lines), and resulting, azimuth integrated profile. Model $A$ has been used for these computations.} 
\label{individual}
\end{figure}

Assume that the observer is situated on the $x$ axis ($x$ axis lies in the plane of the disk) and consider four elements with same radius and with azimuths equal to zero, $\pi/2$, $\pi$ and $3\pi/2$. In the absence of  velocity fields, Stokes $U$ from the first and third element would exactly cancel as well as the Stokes $U$ signal from the second and fourth element. Now, consider the rotating disk. The radiation from the first and third element suffers no redshift and the anisotropy cancels again. However, the radiation from the element no.\,2 is, for example, blueshifted while the radiation from the element no.\,4 is redshifted. As a result, Stokes $U$ profiles do not cancel but result in a double lobed profile as seen at the lower panel of Fig.\,\ref{profiles_model1_static}. It slightly resembles stokes V profiles seen in solar physics and, in some distant sense it can be understood in a similar way: as a superposition of two profiles of the opposite sign, one shifted toward the blue, the other toward the red. This effect would manifest as a rotation of the plane of polarization. For this particular example, the rotation would be rather small as Stokes $Q$ is two orders of magnitude larger than Stokes $U$, however, see the next section for the results for \textbf{externally} illuminated disks. \textbf{To clarify this effect further we show a plot of Stokes $U/I_c$ for different azimuths and given $r \approx r_{\rm{total}}$ (Fig.\,\ref{individual}. It is clearly seen that profiles with opposite Doppler shifts have opposite signs and that they add to an antisymmetric Stokes profile. Furthermore, it is obvious that, at the very disk edge, Stokes $U$ is rather strong (due to large axial anisotropy of the raidation) but it contributes very little to the total polarization.} 

\begin{figure*}
\includegraphics[width = 17cm]{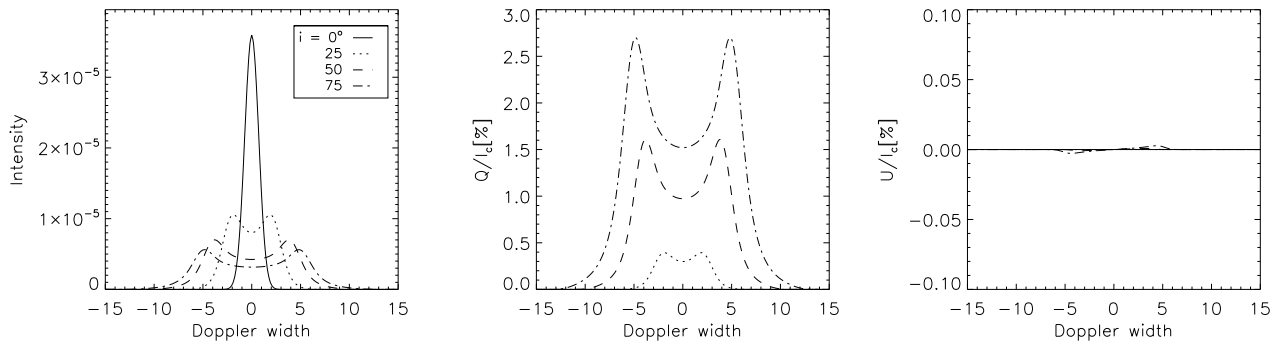}
\caption{Same as Fig.\,\ref{profiles_model1_static}, but for a 100 times less opaque cylinder.}
\label{thin}
\end{figure*}

\textbf{This effect is similar to the ``optical depth effect'' reported by \citet{Vink05} for the case of continuum scattering of frequency-dependent radiation on  the rotating disk. In the case of an optically thin disk  ($\tau \ll 1$), the stokes $U/I_c$ component is supposed to vanish. We illustrate this in  Fig.\,\ref{thin} where we plot the emergent profiles for a rotating disk which is 100 times less opaque than  model A. We plot these profiles only to show the similarity between our interpretation of the presence of the Stokes $U$ component due to the rotation and the interpretation of \citet{Vink05} and therefore we do not use special label for this optically thin model. }

\begin{figure*}
\includegraphics[width = 17cm]{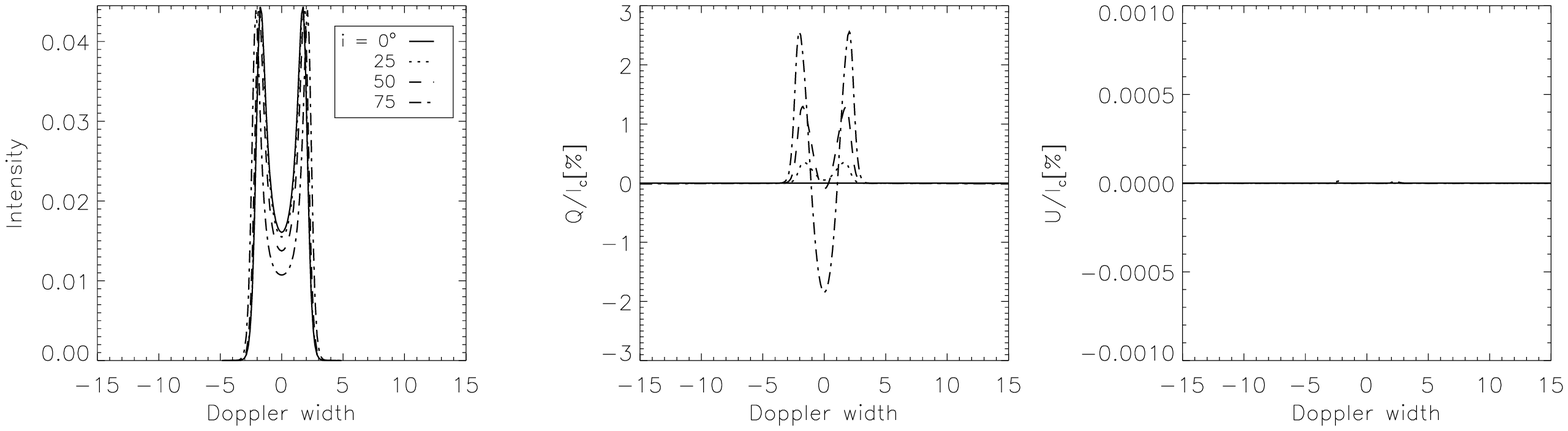}
\includegraphics[width = 17cm]{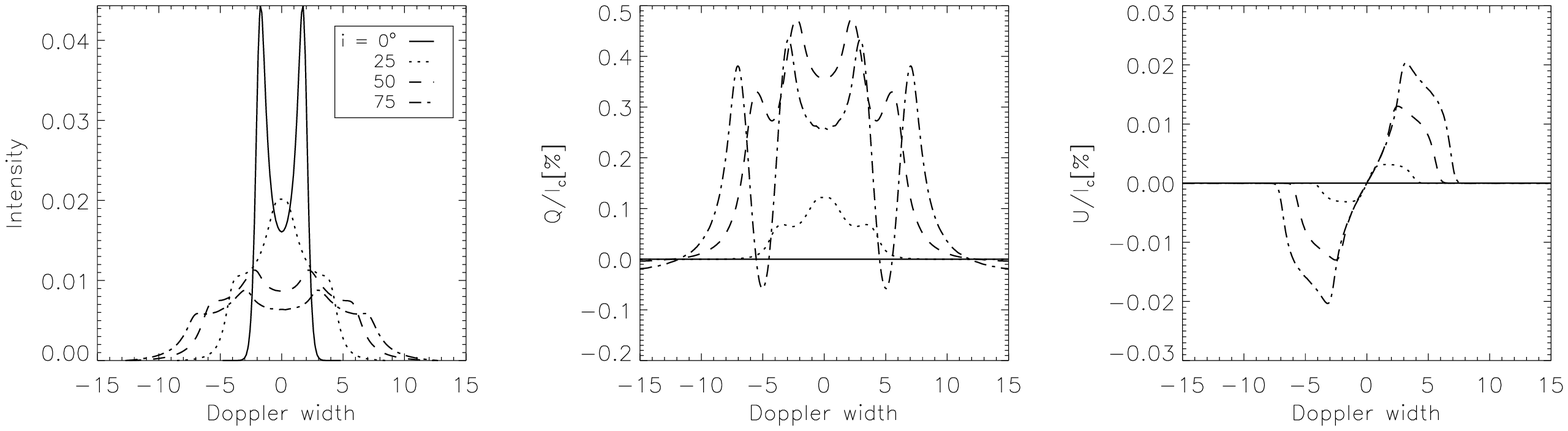}
\caption{Same as Fig.\,\ref{profiles_model1_static}, but for a 10 times more opaque cylinder (Model B). Note that the presence of  rotation brings a significant decrease of Stokes $Q/I_c$.}
\label{profiles_model2}
\end{figure*}

Fig.\,\ref{profiles_model2} shows the emergent profiles from the more opaque model. Intensity profiles now exhibit a deeper ``dip''. This is the consequence of the fact that the near wings (i.e. ``peaks'') are now formed at a depth where the source function is higher. Larger \textbf{values} of the source function at depth is the consequence of the much higher opacity which leads to more successful ``trapping'' of the photons inside the object. $Q/I_c$ polarization in the static case shows profiles similar to the  ones found by \citet{Faurobert88} for the case of semi-infite stellar atmosphere. 

The rotating case is, not surprisingly, more complicated. The maximum intensity in Stokes I profiles decreases and they become broader, even exhibiting hints of \textbf{double} peaks. Although these variations in the core of the line are unlikely to be observed, due to the limited spectral resolution of todays' instruments, broadening is an expected property of profiles formed in rotating objects. An extremely important difference between the rotating and non-rotating cases is that the profiles formed in the static medium are, \emph{without exception} emission profiles with a self-absorption ``dip'' in the center of the line and two peaks, corresponding to the maximum of the emission \citep[for a detailed discussion on the location of these peaks, see][]{EAC12}. Only for a very small values of $v\sin i$ these dips disappear and the profiles resemble simple, one peaked, broadened (although definitely non-Gaussian) emission profile with fine variations, which are, again, likely to be smeared due to the limited spectral resolution.

Scattering polarization line profiles are even more interesting. \textbf{In addition to} (expected) line broadening $Q/I_c$ also shows a significant decrease  at line center. Our interpretation is that this decrease is because, on average, deeper regions of the disk are probed, due to the Doppler shift. Deeper in the medium,  the anisotropy is lower, which means that one is effectively ``seeing'' lower value of  $S_Q$. Stokes $U$ is again present, and increases with $v_{\rm{rot}} \sin i$, due to the same effect as in the case of optically thinner cylinders. However, the values of Stokes U in this example are lower. The explanation for this is the following: The cylinder is more optically thick along $r$, which leads to more isotropic radiation along the azimuth. Ultimately, this leads to lower values of $S_U$ with respect to $S_I$ and to lower signal in Stokes $U/I$. 

To conclude this section we want to emphasize that although these models are extremely simple (homogeneous, isothermal, self-emitting), they clearly demonstrate that the combined effect of line scattering, multidimensional radiative transfer and Keplerian rotation, results not only in a change of line width, but also in a change of line shape. The emergent scattering polarization is also significantly affected. The rotation of the disk clearly results in the presence of Stokes $U$, which changes sign at line center. Furthermore, for disks with  large optical thickness,  rotational velocity also results in a decrease of $Q/I_c$ scattering polarization, apart from the expected broadening of the line profiles. We now turn to the analysis of illuminated disk models.


\subsection{Illuminated disks}

Without doubt, most of gaseous disks do not radiate exclusively by themselves (at least not in optical part of the spectrum, the \textbf{band} we are interested in). In some cases the disk is heated by the central body and it emits thermal radiation so, in radiative transfer terms we can consider it as an emitting body. In other situations, the disk is simply scattering the emergent radiation from the  host object (for example a star). In this section we consider the same models as in the previous one but with with an additional internal illumination. The disks now have a ``hole'' equal to $\frac{1}{5}$ of the disk radial thickness, which hosts a star. To stress again: the models are very simple, homogeneous, isothermal, with radiative transfer parameters given at the beginning of the previous section. The characteristic rotational velocity is, this time,  the velocity on the inner side of the disk, which should be same as the \textbf{break-up speed for the star}. Apart from a Keplerian disk, we also consider one which rotates as a solid body, with angular velocity equal to the inner angular velocity of the Keplerian one. The star in the center is assumed to be spherical in shape and to emit isotropically (i.e. there is no limb darkening). The intensity of the incident radiation is identical to the value of the Planck function in the disk, i.e. equal to unity. The total intensity of the star would in this units be around $0.1$, which means that the disk and the star have approximately the same brightness.

\textbf{In the subsequent results, we do not model any occultation effects. One occultation effect is the star occulting the disk and thus breaking the axial symmetry. The modeling of this effect is very important as it can alter the shape of the Stokes $U$ \citep[as is shown in, for example,][]{Milic13}. We chose not to include it here as we are already dealing with several different influences to the polarized line profile shape. However, we will include it in our further work which will involve more realistic disk models. The other effect is the occultation of the star \emph{by} the disk. We do not add the radiation emerging from the un-occulted stellar surface to the flux. This effect cannot change the shape of the lines but can just alter the continuum level, and increase the amount of unpolarized flux, thus reducing the degree of polarization of the light. We will include this effect in our upcoming work as well.}



\begin{figure*}
\centering
\includegraphics[width = 17cm]{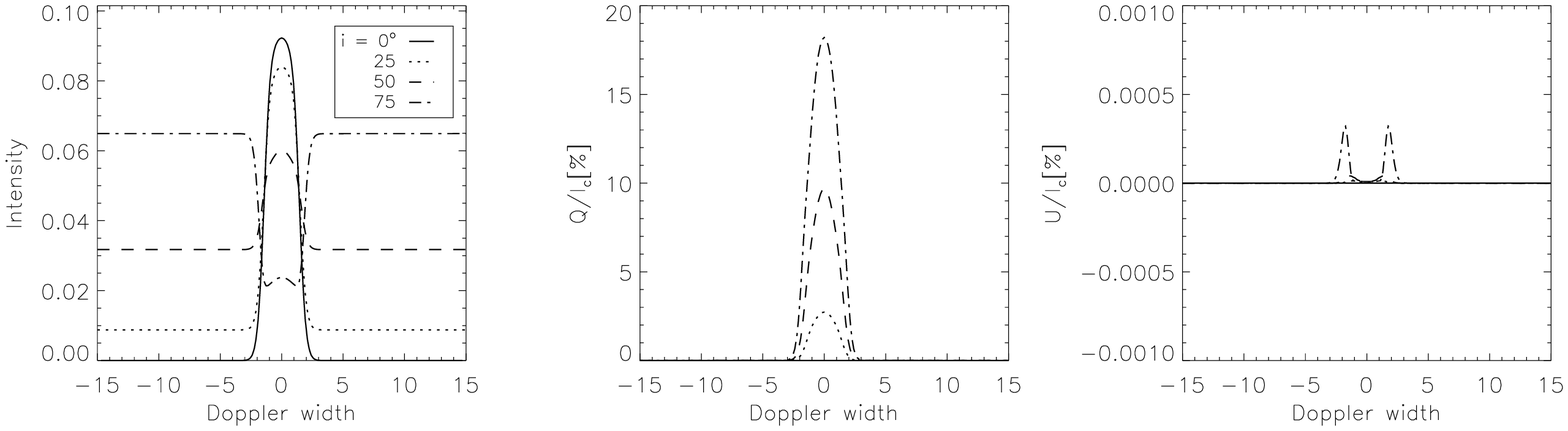}
\includegraphics[width = 17cm]{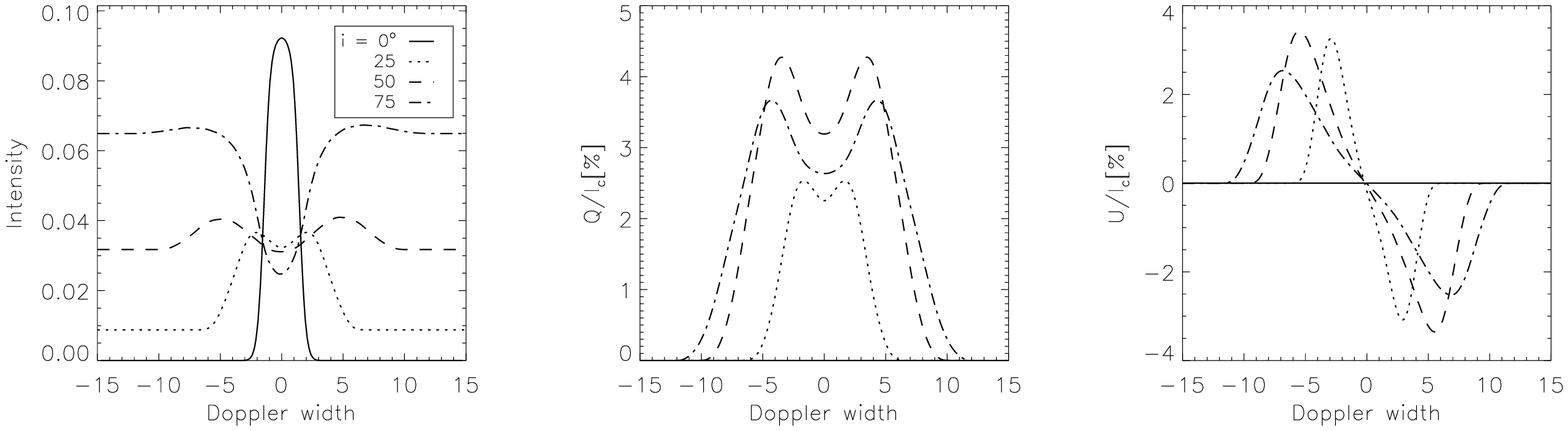}
\includegraphics[width = 17cm]{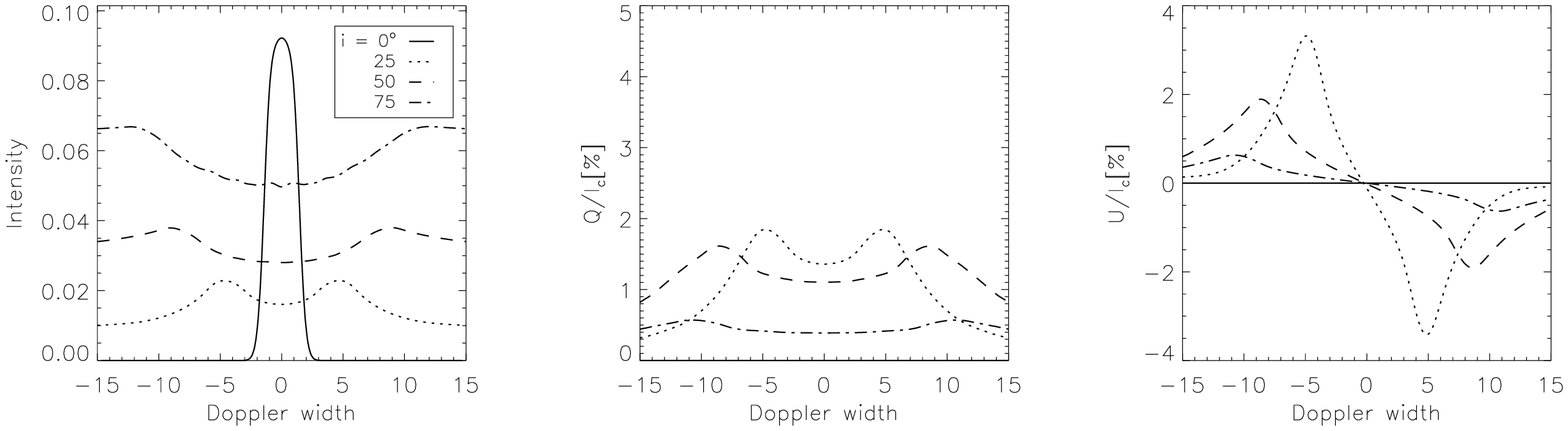}
\caption{Profiles emerging from the disk illuminated from the inside. The disk has a radial optical thickness $\tau_r$ = 1000 and a vertical optical thickness $\tau_z$=10 (Model A$_i$). Top: non-rotating disk; middle: Keplerian rotation; bottom: solid body rotation. Note that the intensity profile for the highest inclination has been scaled by a factor of 0.5. See the text for the discussion of line profile shapes.}
\label{profiles_model3}
\end{figure*}

Fig.\,\ref{profiles_model3} shows the emergent profile from the less opaque disk model ($\tau_z = 10$, $\tau_r=1000$) for the non rotating and rotating case. First, note the presence of the continuum due to the transmitted light of the star in the far wings of the line. In the presence of the illuminating object (which means that the inner boundary condition for the intensity is non-zero), the emitted radiation of the disk consists of three parts: i) the transmitted and attenuated light from the host star ii) the scattered incident \textbf{stellar} radiation; and iii) the thermally emitted (and perhaps again scattered) radiation of the disk itself. From the first contribution we expect an absorption line, from the other two, an emission one (as the object is not infinitely optically thick). We see, indeed, that  when the inclination increases the emission is getting weaker and weaker with respect to the transmitted continuum and ultimately the line turns into an absorption line with some emission in the line center. Scattering polarization in $Q/I_c$ is, as expected, increasing with increasing inclination and is much higher than in the previous case, predominantly because of the anisotropy introduced by the presence of the source of illumination. Stokes $U$ is again essentially equal to zero.

The presence of  rotation changes the observed profiles a lot. Stokes $I$ is now broader and ``flatter'' and we find it very possible that for specific range of inclinations the line disappears as the emission and absorption contributions exactly cancel. Interestingly, the scattering polarization would remain, as the indicator of the presence of the disk! Stokes $Q/I_c$ is not monotonic with respect to the inclination any more because of the effect of velocity fields similar to the second example described in the previous section. The Stokes $U/I_c$ signal is significant as the anisotropy is much higher, due to the illuminating star, and the ``splitting'' effect due to the rotation remains, producing a double-lobed profile. This is true both for disks which rotate obeying a Keplerian velocity law and for rigidly rotating ones. Generally, line profiles have similar shapes, but the ones emerging from the disk which rotates as a rigid body are much wider (because the velocity is constantly increasing, instead of decreasing, toward the disk edges), and subsequently, the polarization degrees are significantly lower.  

It is of interest to inspect also the variation of the total line polarization $P$ and polarization angle $\Theta$ with respect to the wavelength (and, also, those two quantities are perhaps more often used in stellar physics community). They are defined as:
\begin{equation}
P = \frac{(Q^2 + U^2)^{1/2}}{I}
\end{equation}
and:
\begin{equation}
\Theta = \frac{1}{2} \tan^{-1}\frac{U}{Q}
\end{equation}

\begin{figure*}
\sidecaption
\includegraphics[width = 12cm]{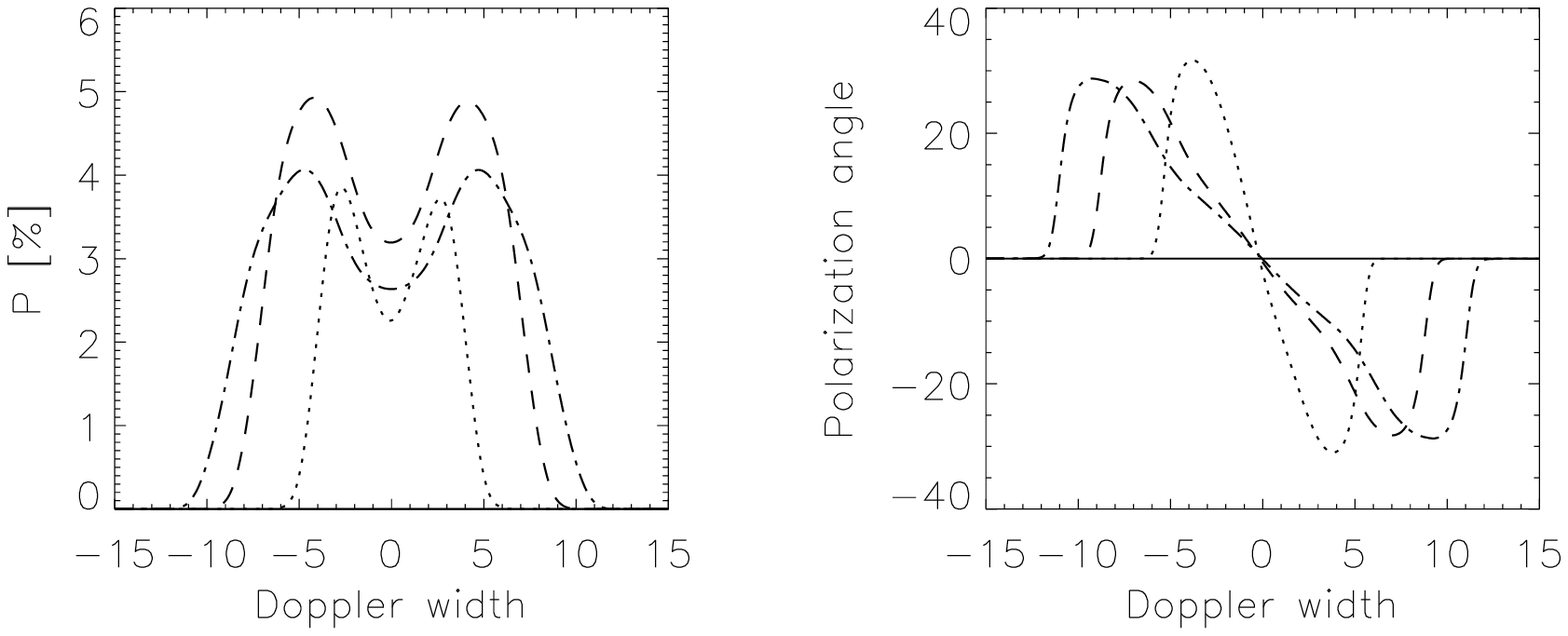}\\
\includegraphics[width = 12cm]{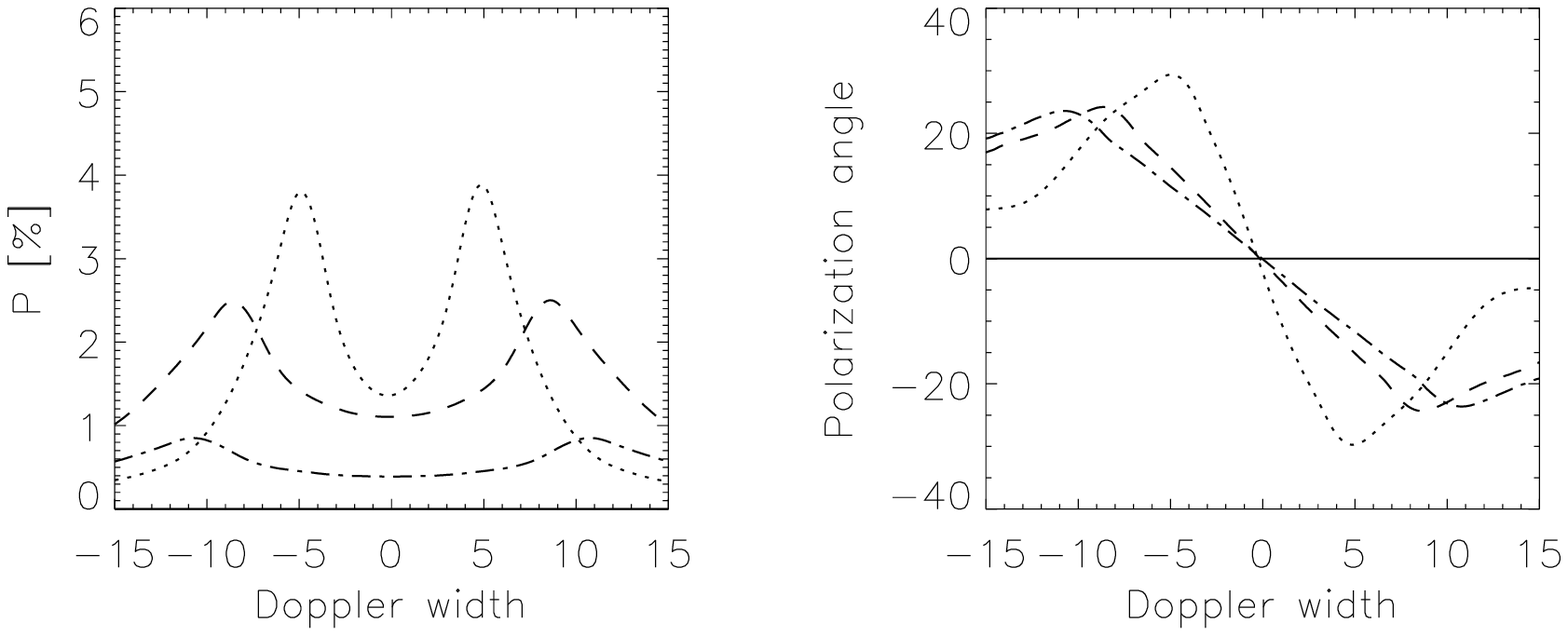}
\caption{Variation of total linear polarization $P$ and polarization angle $\Theta$ along the line for the internally illuminated homogeneous disk with radial optical thickness equal to $1000$ and horizontal optical thickness equal to $10$ (model A$_i$). Top: Keplerian rotation; bottom: solid body rotation.}
\label{model3_alternative}
\end{figure*}

\begin{figure}
\includegraphics[width = 0.5\textwidth]{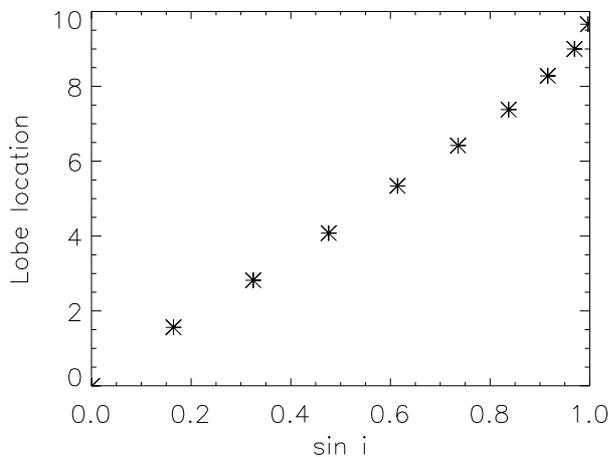}
\caption{Dependance of half-spacing between maximum of the lobes in the polarization angle line profile on the intensity for the model $A_i$.}
\label{vsini}
\end{figure}

Fig.\,\ref{model3_alternative} shows that the total polarization in the line, for a Keperian disk, changes non-monotonically with inclination. This is the effect sometimes known as the ``dilution of the polarization by unpolarized radiation'': Part of the original, unpolarized, flux of the star is transmitted and is effectively reducing the total polarization for high inclinations. Again, we want to stress that this is the case where we study only the radiation coming from the disk,  the situation is likely to change once we introduce properly the obscuration effect by the host star (this will be done in  forthcoming papers). It is interesting that the maximum polarization angle does not change much with inclination, but the spacing between lobes does. This effect of the change of the sign of polarization angle has been studied in more detail by \citet{Vink05}, although for the case of coherent scattering of line radiation on free electrons. It would be extremely interesting to study the differences in this effect between coherent and resonant line scattering and see if the type of scattering leaves imprint in the variation of the polarization angle across the line. Anyway, this effect could be very interesting in better constraining the rotational velocities of disks. To illustrate that, we plot the half-spacing between the two rotation angle lobes versus $\sin\,i$ (Fig.\ref{vsini}). It can be seen that the dependance obeys almost a perfect linear law with a slope equal to the rotational velocity at the inner disk boundary. This means, in principle, that measurements of polarization angle in spectral lines could be used to deduce the value of $v\,\sin{i}$ of circumstellar disks.

The effects are similar for disks obeying a solid body rotation, except that the total polarization is, for this particular case, actually \emph{decreasing} with inclination. Here we obviously witness the competition between increasing scattering angle (as $i$ approaches $\pi/2$ the polarization increases) and increased broadening (broader lines result in smaller maximum polarization). An interesting aspect is that, while the lobes where the polarization angles reach minimum/maximum, have approximately the same location in Keplerian and rigidly rotating case, in the latter we remark that the polarization angle decreases very slowly in the line wings. 

\begin{figure*}
\centering
\includegraphics[width = 17cm]{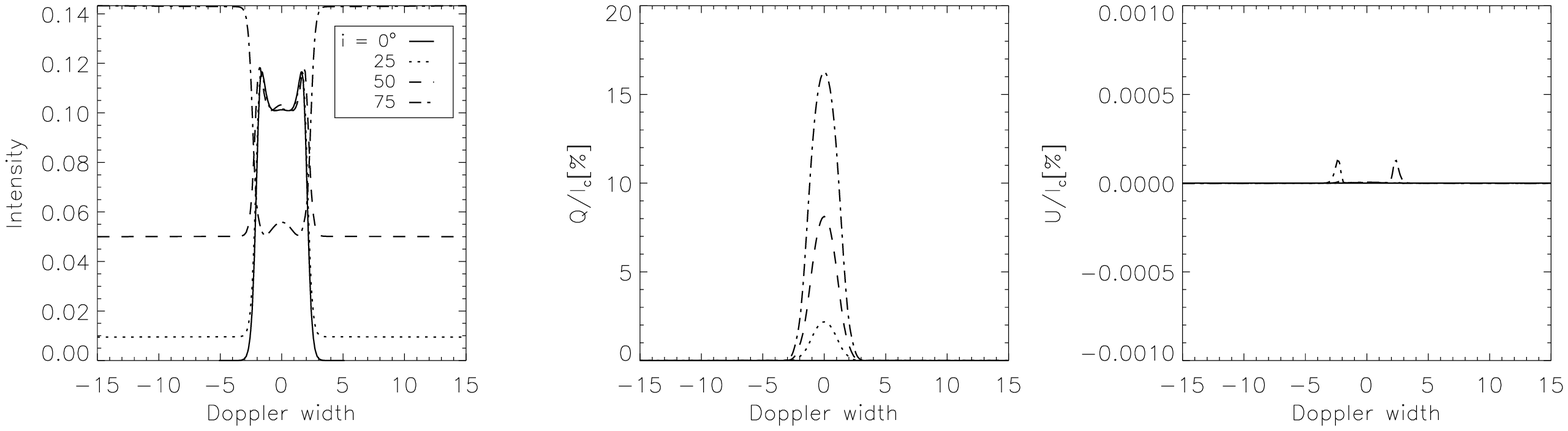}
\includegraphics[width = 17cm]{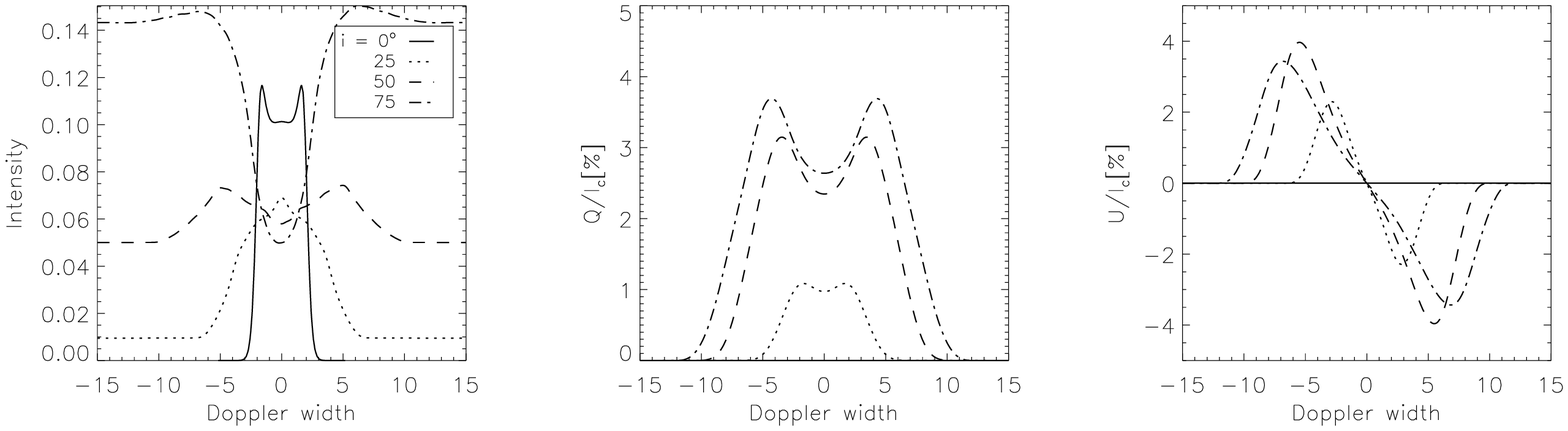}
\includegraphics[width = 17cm]{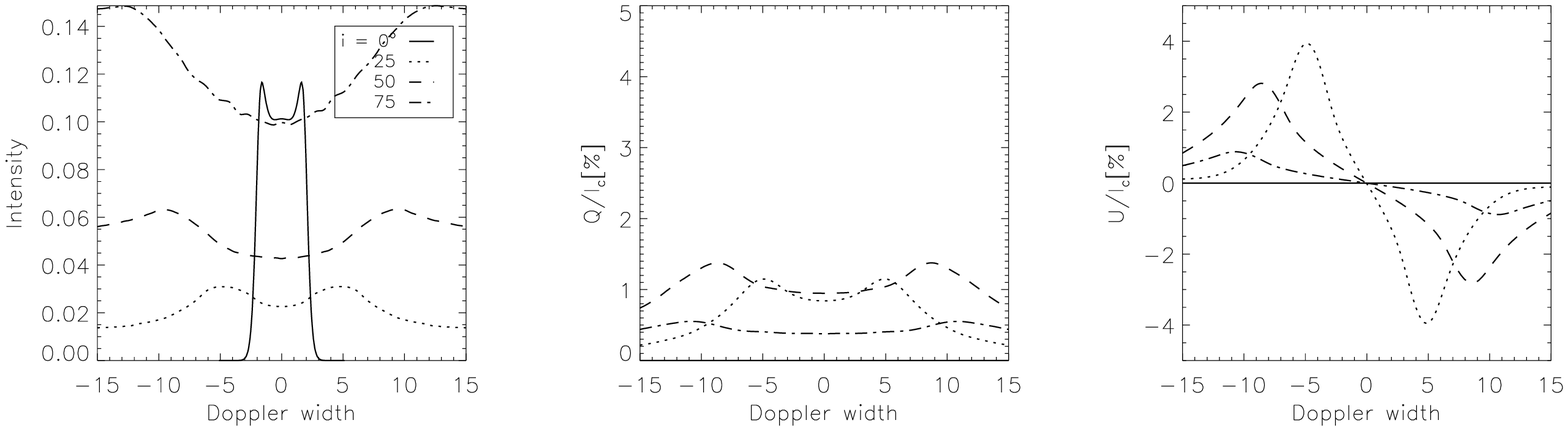}
\caption{Same as Fig.\,\ref{profiles_model3}, but for 10 more opaque disk (model B$_i$).}
\label{profiles_model4}
\end{figure*}

To analyze further the effects of optical thickness and NLTE radiative transfer (i.e. multiple resonant line scattering), we again re-do the computations but for 10 times more optically thick disk (Model B$_i$). In such a case, the radiation coming from the disk itself is non-negligible and it contributes mostly to the intensity profiles where, for the lower three inclinations we see a self-absorbing, emission line, superimposed on a background (transmitted) continuum (Fig.\,\ref{profiles_model4}). The linear polarization in Stokes $Q$ is as high as in the optically thinner case, while Stokes $U$ is negligible, as expected.

In the presence of rotation, we see the polarized profiles very similar to the optically thinner disk (Fig.\,\ref{profiles_model3}). Obviously the polarization is formed in the region where the polarized source function is dominated by scattered radiation from the incident star. One should keep in mind that the degree of polarization depends only on the anisotropy which is, obviously dominated by the anisotropic illumination of the incident radiation. This, in turn, results in polarized profiles which are \textbf{relatively} insensitive to the opacity of the disk.

\begin{figure*}
\sidecaption
\includegraphics[width = 12cm]{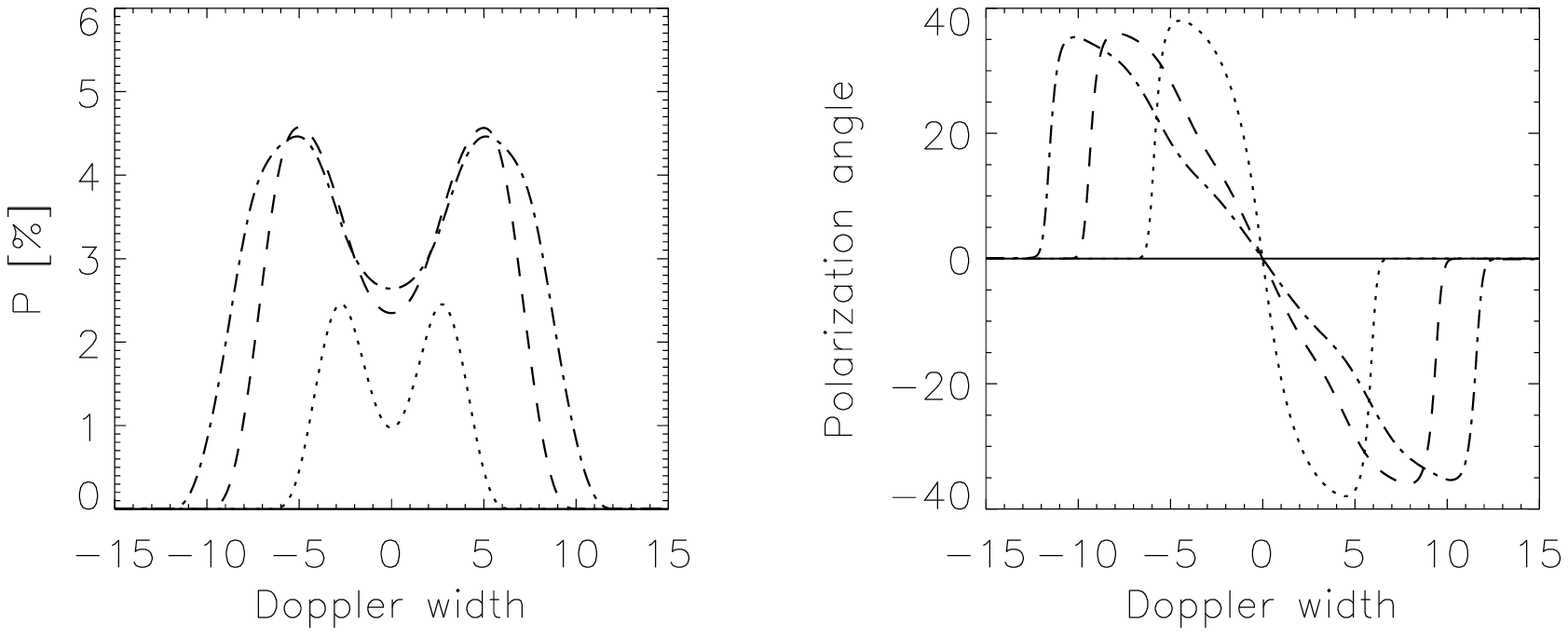}\\
\includegraphics[width = 12cm]{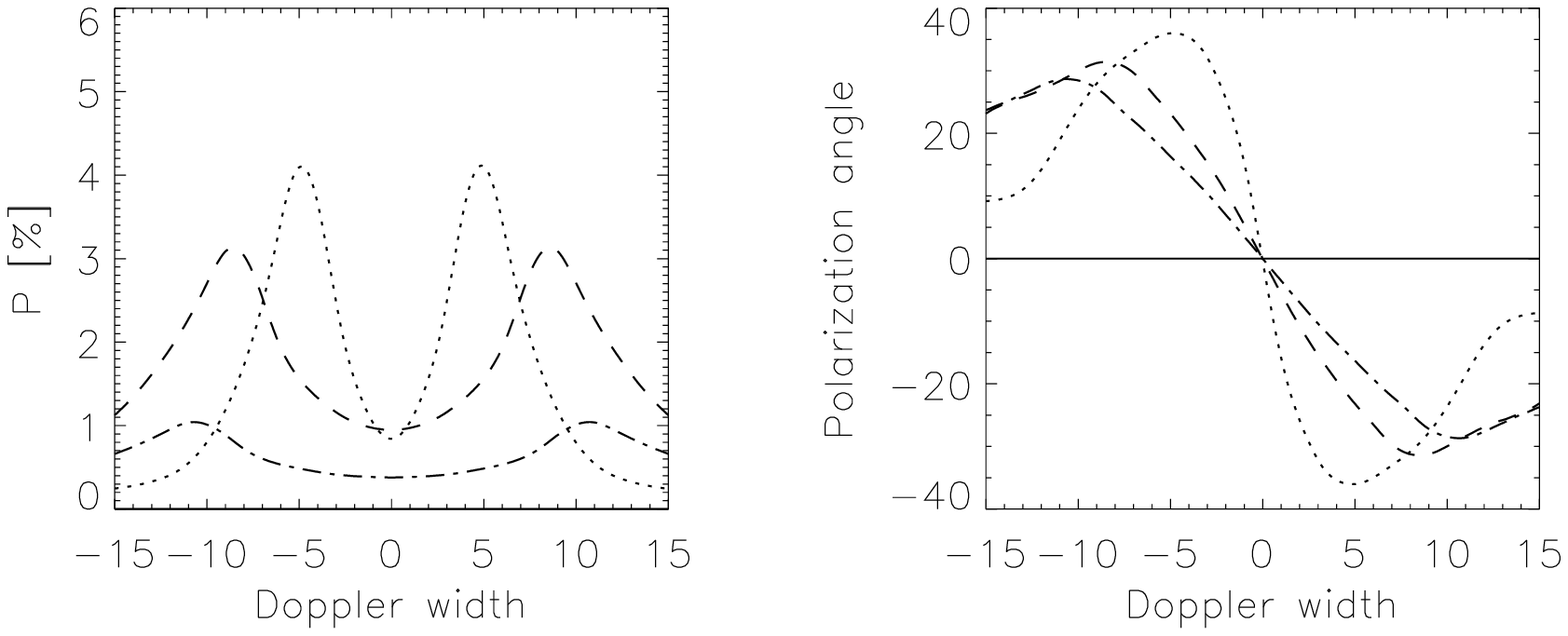}
\caption{Same as figure\,\ref{model3_alternative} but for 10 times more opaque disk (model B$_1$).}
\label{model4_alternative}
\end{figure*}

 Only for the low inclination case ($i=25^o$) we see a drop in the degree of linear polarization both in stokes $Q$ and $U$. Our interpretation is that this is due to the small scattering angle combined with the effect of multiple scatterings which destroy the polarization. For higher inclinations, on the other side, both optically thinner and thicker disks exhibit similar polarized profiles, which is interesting. \textbf{We also show the total polarization in the line and variation of the polarization angle over the line for this more opaque disk model (Fig.\,\ref{model4_alternative}).}

Also, the variation of the polarization angle across the line is very similar to the previous case (model A$_i$), which is good news because it means that the potential diagnostic technique\footnote{Such a diagnostic has already been presented by \citet{Smith05}, and recently used by \citet{Luka14}, but for AGN disks, not circumstellar ones.} which relies upon this is not sensitive to the optical depth of the disk in the observed spectral line. Probably the most interesting result is that, contrary to the case of self-emitting disks, an increase of the optical thickness of the disk does not change significantly the degree of the polarization.  

\section{Conclusions}

In this paper we have examined the effect of rotation on emergent line scattering polarization in simplified models of gaseous disks. We have considered both disks which are self-emitting and ones illuminated by an inside source (i.e. star). The models are very simple: homogeneous, isothermal and rotating obeying Keplerian law. To compute the emergent polarized line profiles we have used a fully consistent way of solving the equations of radiative transfer and statistical equilibrium for a two level atom in the presence of velocity fields. We have employed the so-called reduced intensity formalism, which has been so far used in solar physics but not for the modeling of other objects. We have solved the radiative transfer equation in 2D cylindrical geometry which is not commonly used, but, for this case, is an advantageous approach.

Even in these, perhaps unrealistic models, the parameter space to explore is rather large: disk optical thickness, rotation velocity, size of the star with respect to the disk, ratio between disk radial and horizontal size... Therefore, it is rather hard to \emph{investigate} the effects of different parameters. Needless to say, in realistic models, also temperature, opacity and photon destruction probability can vary with spatial coordinates and rotation can deviate from the Keplerian law. That is why, in this paper, we have restricted ourselves to attempt to \emph{demonstrate} the effects of rotation on emergent polarized line profiles. In this paper we have considered rather low rotational velocities ($v_{\rm{max}} \approx 10 v_D$), but the presence of these, even relatively weak velocity fields, causes significant changes in the polarized line profiles. For internally illuminated disks, we have also computed polarized profiles for disk obeying solid body rotation, to illustrate the differences between different rotational laws.


In the case of self-emitting disks, the main effects of rotation are the broadening of Stokes $I$ and $Q/I_c$ profiles and the presence of weak linear polarization in Stokes $U/I_c$. In the optically thinner case ($\tau_r = 1000$, $\tau_z = 10$), the $Q/I_c$ polarization at line center does not change significantly when a rotation of the disk is introduced, while in the optically thicker case (opacity  10 times higher), the presence of rotation decreases the degree of linear polarization by about a factor of 5. The reason for this is, in our interpretation, \textbf{the different behavior} of the radiation anisotropy  in the disk between the two cases. We stress once again that in neither of these cases is the disk optically thin (that is optical thickness along each axis is much greater than unity) nor optically thick (i.e. the size of the disk along at least one axis is smaller than the photon thermalization length, $1/\epsilon$). In solar/stellar atmosphere modeling it is customary to refer to such a medium as ``effectively thin'', and this is exactly the case which is most interesting for radiative transfer investigations in multidimensional objects. 

Internally illuminated disks show more interesting features. Due to the presence of the anisotropic illuminating radiation, the degree of polarization is much higher. Subsequently, the ``spectral splitting'' between the receding (red shifted) and approaching (blue shifted) parts of the rotating disk is much more prominent in Stokes $U/I_c$ which now exhibits 
polarization degrees of a few percent. Talking in terms of angle of polarization ($\Theta = \frac{1}{2} \tan^{-1} \frac{U}{Q}$), this results in a significant variation of the polarization angle across the line. Interestingly, this angle does not seem to be sensitive neither to the optical thickness of the disk, nor to the inclination (actually a combination of the rotational velocity and inclination, $v\,\sin i$). This means that this effect can be used for disk diagnostics, which has already been pointed out by \citet{Vink05} and \citet{Smith05} for the case of coherent scattering. 

To conclude this paper we want to emphasize that the presence of rotation tremendously influences the polarized line formation process in disk-like objects. Large influence opens up space for devising intelligent diagnostic techniques, which is the scope of further research. As spatially resolved spectropolarimetric observations of such objects are still unfeasible, one is to devise diagnostic techniques which rely on spatially integrated spectropolarimetric observations. In such studies, detailed line formation modeling such as one hinted in this paper might prove to be necessary and we think that techniques routinely used in solar and stellar physics should be extended to line formation in disk-like objects as well. Probably the greatest insight into the physics of these objects can be obtained by using state-of-the-art spectropolatimetric (ESPaDOnS, NARVAL) together with spectro-interferometric (CHARA, VEGA) observations. We hope that modeling approaches such as the ones presented in this paper can help in the interpretation of such high-quality observations.

Our future work will present results from more realistic disk models where opacity and temperature vary inside the object, and also first attempts to model observed spectral features. Also, we aim to introduce the so called co-moving frame formalism for the solution of radiative transfer equation in  moving media, in order to be able to model rapidly rotating objects such as different kinds of accretion disks. 

\begin{acknowledgements}

\textbf{We are indebted to the anonymous referee for the detailed review which significantly improved quality of the manuscript.} We also thank Jorrick Vink, Slobodan Jankov and Olga Atanackovi\'{c} for useful comments and critical reading of the manuscript.

IM is grateful to Pavle Savi\'{c} grant, COST action 1104 and SOLARNET project for financially supporting his research and collaboration with MF. This research is also partly funded by Serbian Ministry of Science and Education, under the project 176004, ``Stellar physics.''

\end{acknowledgements}

\appendix
\section{Approximate calculations of the emergent Stokes profiles from a rotating disk}

To further illustrate the differences (or lack thereof) between different approaches to the computation of the formal solution, we present here emergent Stokes profiles from model A computed with three different approaches:
\begin{enumerate}
\item We compute the reduced source function in the presence of the velocity fields and then use it to formally solve the RTE in moving media. This approach is without any approximations and only errors arise due to numerical inaccuracies. 
\item We use the source function computed in the static case and formally solve the RTE in the moving media. Formal solution here is completely consistent while the source function is approximated to be identical to the static case.
\item We use the source function computed in the static case and formally solve the RTE in the \emph{static media}. We then appropriately Doppler-shift the profiles in the process of spatial (that is, angular) integration (Eq.\,\ref{integration_app}). In this approximation we assume that each ``slice'' of the disk behaves like a static object in terms of radiative transfer and that all changes to the spectrum arise purely because of the Doppler shift of individual disk parts (i.e. there is no lateral radiative transfer). 
\end{enumerate}

\begin{figure*}
\centering
\includegraphics[width = 17cm]{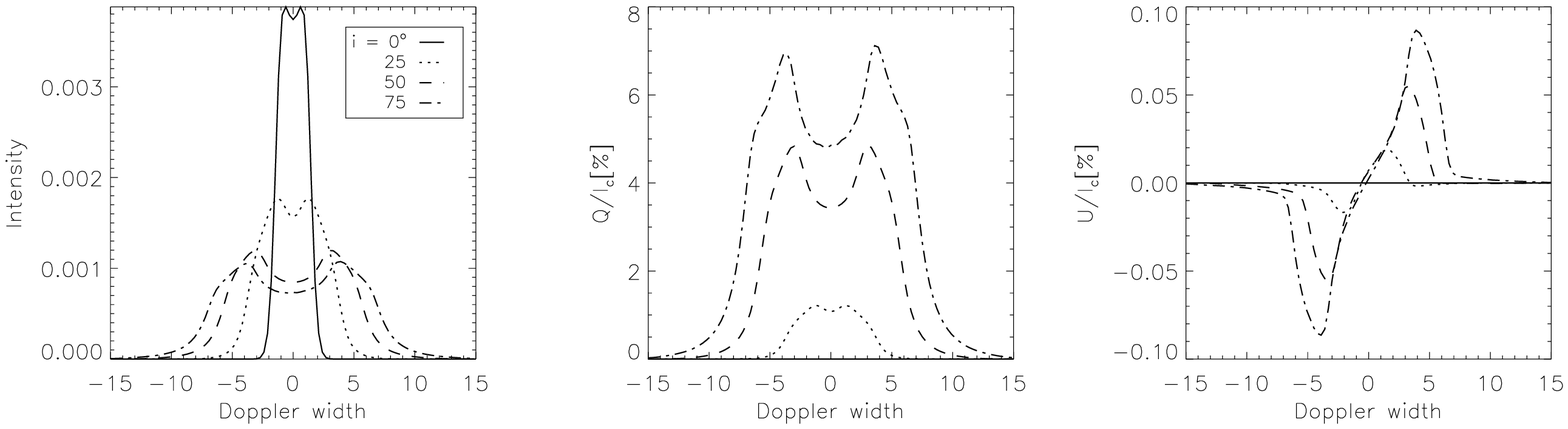}
\includegraphics[width = 17cm]{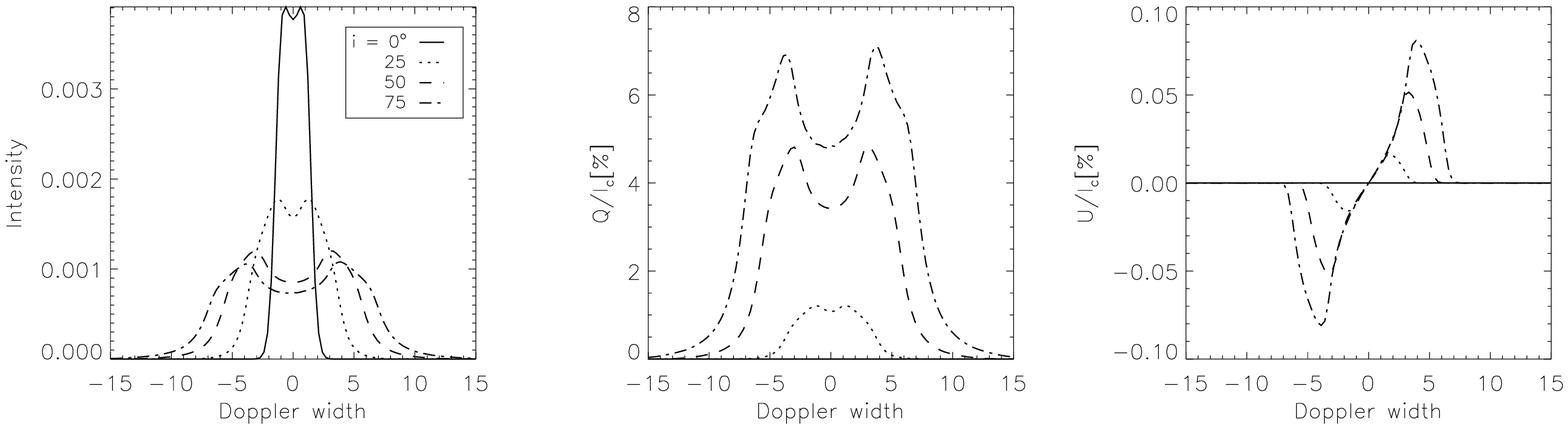}
\includegraphics[width = 17cm]{profiles_model1_v=5.eps}
\caption{Comparison between polarized line profiles computed from the model A in the presence of Keplerian rotation using approaches 1 (upper panel), 2 (middle panel) and 3(lower panel) described in the appendix.}
\label{comparison}
\end{figure*}

Emergent profiles are shown in the Fig.\,\ref{comparison}. As the whole paper is focused on a qualitative analysis of the line shapes we restrict ourselves to visual inspection of the differences. To justify this, we remind the reader that the current state of spectropolarimetric observations of objects outside of the solar system are rather inaccurate \citep[for example, see ][for some observations of Herbig Ae/Be stars]{Vink02} concerning the polarimetric aspect. 

The notable difference is  the presence of a very weak Stokes U in the far wings on the line which can be seen only when the approach 1 is used. However, as this effect is more than an order of magnitude weaker than the maximum polarization in Stokes U we feel that the difference with respect to two other cases is negligible. In the top two plots also a slight asymmetry in Stokes $Q/I_c$ can be seen. 
This is the effect of an insufficient number of points in $\phi$, which results in numerical inaccuracies. As is stated in the text, the integration over the azimuth is actually the \emph{spatial integration}, because we assume that the disk is axisymmetric. For computations by approaches 1 and 2 we have used 36 Gaussian points in Azimuth and 121  equally spaced points in frequency. Note that the major advantage of the approach 3 is that the number of frequency points can be significantly smaller as there is no need to cover all possible red/blue shifts in the frequency mesh.

\bibliographystyle{aa}  
\bibliography{ivan_disks} 

\end{document}